\documentstyle[12pt]{article}
\font\msbm=msbm10 scaled 1000
\newfam\msbmfam
\textfont\msbmfam=\msbm
\def\Bbb{\fam\msbmfam\msbm}
\font\scs=cmcsc10       

\newtheorem{prop}{Proposition}[section]
\newtheorem{theorem}[prop]{Theorem}
\begin{document}
\normalbaselineskip=40pt

\title{Connection formulae\\
for the degenerated asymptotic solutions\\
of the fourth Painlev\'e equation}
\author{Andrei A. Kapaev}
\date{}
\maketitle

\vskip1.5cm
\centerline{\it St.~Petersburg Department of Steklov Mathematical Institute}
\centerline{\it of Russian Academy of Sciences,}
\centerline{\it Fontanka 27, St.~Petersburg, 191011, Russia}

\vskip2cm
\centerline{\bf Abstract}
{\small All possible 1-parametric classical and transcendent
degenerated solutions of the fourth Painlev\'e equation with the corresponding
connection formulae of the asymptotic parameters are described.}

\newpage
\section{Introduction}\label{introduction}
\vskip 1em
\baselineskip=20pt

We consider the general case of the fourth Painlev\'e equation P4 \cite{I}
\begin{equation}\label{p4}
y''={(y')^2\over 2y}+{3\over 2}y^3+4xy^2+2\bigl(x^2-a\bigr)y-{b\over 2y}\,,
\end{equation}
various physical applications of which are presented in the papers
\cite{ARS}--\cite{SV}.

Among the most important results in the P4 theory obtained without any use
of inverse problem method, we mention the article of Lukashevich \cite{L},
where the basic B\"acklund transformations and rational solutions of P4 are
constructed, the papers of Airault \cite{A}, Gromak and Lukashevich \cite{GL},
who find the Riccati equation for the classical 1-solutions of P4, the works
of Umemura and Watanabe \cite{UW} and Gromak \cite{G} on irreducibility of the
general fourth Painlev\'e transcendent. Survey of these and many other results
is given in the paper \cite{BCH}.

The relation of the Painlev\'e equation theory with the problem of the
isomonodromy deformations of the Fuchsian equations discovered in classical
works of Fuchs \cite{F}, Schlesinger \cite{Sch} and Garnier \cite{Grn},
allows Flaschka and Newell \cite{FN}, Jimbo and Miwa \cite{JM} and Its and
Novokshenov \cite{IN} to develop an inverse problem method for investigation
of these equations, which is applied
to P4 in articles of Fokas, Mugan, Ablowitz and Zhou \cite{FMA, FZ}. The
authors parameterize the fourth Painlev\'e transcendent set via the
monodromy data of the associated linear system of Jimbo and Miwa \cite{JM},
find some B\"acklund transformations,
describe the proper Riemann-Hilbert problem as some factorization problem for
a piece-wise holomorphic function and reduce it to some system of integral
Fredholm equations. The last allows them to prove both the Riemann-Hilbert
problem solvability in the case of general position and the classical
theorem of meromorphy of the Painlev\'e function of the fourth kind.

Using the fact revealed by Ablowitz, Ramani and Segur \cite{ARS} that P4
describes the similarity reduction of the Derivative Nonlinear Schr\"odinger
(DNS) equation, Kitaev \cite{Kit(1985)} finds an alternative to \cite{JM} Lax
pair by use of reduction of the Lax pair for DNS of Kaup and Newell \cite{KN}.
Later, in \cite{Kit(1988)}, he describes the Schlesinger transformations of
this linear system and the corresponding B\"acklund transformations of the
fourth Painlev\'e transcendent, one of them coincides with the transformation
found by Lukashevich \cite{L} and other looks new but, as it is shown in
\cite{BCH}, is a superposition of the transformations of Lukashevich. Later,
in ref.\ \cite{MCB}, Milne, Clarkson and Bassom repeat some of the
calculations of Kitaev and evaluate the alternative Lax pair (in less
symmetric form, though) and one of the B\"acklund transformations for P4. They
also find the monodromy data for some simplest classical solutions of P4.
In the work \cite{Kap}, the case of general 2-parametric Painlev\'e function
is described asymptotically by use of isomonodromy method for any $\arg x$,
$|x|\to\infty$, including elliptic and trigonometric asymptotic solutions
with the corresponding connection formulae.

Bassom, Clarkson, Hicks and McLeod, in their paper \cite{BCHM}, begin
intensive
investigation of the classical solutions of the fourth Painlev\'e equation.
In particular, they get the connection formulae for some classical solutions
of the nonlinear harmonic oscillator related to P4 (\ref{p4}) with $b=0$ and
integer $a$. Some ideas of the paper are developed further in ref.\
\cite{BCH}, where huge tables of exact solutions (rational and classical) of
P4 accompanied by the numerical calculations and pictures are presented. All
the solutions are obtained by applying the B\"acklund transformations to some
``seed" rational and classical solutions. However, all these explicit results
are almost useless for asymptotic investigation of the classical solutions
and the connection formulae evaluation.

The classical solutions expressible in terms of the parabolic cylinder
functions are the special cases of the so-called degenerated (in other terms,
instanton, separatrix or truncated) solutions. Mostly, these solutions have
the form of some asymptotically algebraic background satisfying the
Painlev\'e equation with some additive exponentially small term
depending on the initial data. Besides the classical solutions, there exist
other degenerated Painlev\'e transcendents. In ref.\ \cite{Kap}, some of them
are described as limiting cases of the transcendent 2-parametric solutions of
general positions. Other transcendent degenerated solution of P4 ($b=0$)
exponentially decreasing at $+\infty$ is investigated by Its and Kapaev in
ref.\ \cite{IK} via the Riemann-Hilbert problem method.

The main idea of the present paper consists in application of the B\"acklund
transformations to the formal asymptotic solutions. Since the set of
degenerated asymptotics is invariant under action of the B\"acklund
transformation group and there is a finite number of essentially different
degenerated asymptotic ansatzes, the action of any B\"acklund transformation
can be effectively described as the superposition of permutation of the
asymptotic ansatzes and some change of the asymptotic parameters involved.
This procedure allows us to show that the simplest of the classical solutions
satisfying the Riccati equations, the decreasing asymptotic solution from
\cite{IK} and the degenerated asymptotics found in \cite{Kap}, affected by
the B\"acklund transformation chain, yield all the degenerated solutions of
the fourth Painlev\'e equation and to get the complete description of any
degenerated Painlev\'e function with the connection formulae.

In the section~\ref{lax_pair}, we describe the fourth Painlev\'e equation as
the equation of the isomonodromy deformation for the alternative linear
differential system found by Kitaev and give the proper parametrization of
the Painlev\'e transcendent set via the points of the monodromy manifold.
The section~\ref{backlund_schlesinger} contains description of the group of
the B\"acklund transformations of P4 generated by the group of the Schlesinger
transformations for the corresponding linear system. The next
section~\ref{classical_solutions} gives some information on the simplest
classical solutions of P4, their asymptotics and coordinates on the monodromy
manifold. In the section~\ref{degenerated_solutions}, we consider the set of
the formal asymptotic solutions of P4 of the separatrix class and the action
of the B\"acklund transformation group on this invariant set. Using the
results, we give the complete asymptotic description of the classical
Painlev\'e functions with the connection formulae. In the
section~\ref{connection}, we describe the transcendent degenerated solutions
of P4 with the corresponding connection formulae of the asymptotic parameters
with the monodromy data of the associated linear system.

\section{P4 as a monodromy preserving deformation}\label{lax_pair}

Let us consider the Lax pair for P4 (\ref{p4}) of Kitaev \cite{Kit(1985)},
alternative to the pair found by Jimbo and Miwa \cite{JM}:
\begin{eqnarray}\label{lambda_equation}
&&\frac{\partial\Psi}{\partial\lambda}=
\biggl\{\Bigl(\frac{1}{2}\lambda^3+\lambda(x+uv)+
\frac{\alpha}{\lambda}\Bigr)\sigma_3+
i\Bigl(\lambda^2u+2xu+u'\Bigr)\sigma_++\nonumber
\\[4pt]
&&+i\Bigl(\lambda^2v+2xv-v'\Bigr)\sigma_-\biggr\}\Psi\, ,
\\[4pt]\label{x_equation}
&&\frac{\partial\Psi}{\partial x}=
\biggl\{\Bigl(\frac{1}{2}\lambda^2+uv\Bigr)\sigma_3+
i\lambda u\sigma_++i\lambda v\sigma_-\biggr\}\Psi\, ,
\end{eqnarray}
where
$\Psi$ is some $2\times2$ matrix valued function of the complex variable
$\lambda$ depending on the complex $x$, $u$, $v$, $\alpha$, and
$$
\sigma_3=\pmatrix{1&0\cr 0&-1\cr},\qquad
\sigma_+=\pmatrix{&1\cr 0&\cr},\qquad
\sigma_-=\pmatrix{&0\cr 1&\cr}.$$

Compatibility of the equations (\ref{lambda_equation}), (\ref{x_equation})
implies that $\alpha$ does not depend on $x$, while $u$ and $v$
are the functions of $x$ satisfying the system of differential equations
\begin{eqnarray}\label{system_uv}
&&u''=-(1+2\alpha)u-2xu'+4xu^2v+2u'uv\, ,\nonumber
\\[4pt]
&&v''=(1-2\alpha)v+2xv'+4xuv^2-2uv'v\, .
\end{eqnarray}
In particular, the system (\ref{system_uv}) implies the constancy of the
combination
\begin{equation}\label{beta}
\beta=u'v-uv'+2xuv-(uv)^2\equiv const\,,
\end{equation}
and yields the product
\begin{equation}\label{y}
y=uv
\end{equation}
satisfies the equation P4 (\ref{p4})
\begin{equation}\label{p4ab}
y''={(y')^2\over 2y}+{3\over 2}y^3+4xy^2+
2\Bigl(x^2-2\alpha+\frac{\beta}{2}\Bigr)y-\frac{\beta^2}{2y}\,,
\end{equation}
with the parameters
\begin{equation}\label{ab_ab}
a=2\alpha-\frac{\beta}{2},\quad b=\beta^2\, .
\end{equation}
{\it Remark}: The relation (\ref{ab_ab}) means that each Painlev\'e function
is related to two linear systems (\ref{lambda_equation}) which differ from
each other by the sign of parameter $\beta=\pm\sqrt{b}$ and the corresponding
value of the parameter $\alpha=\frac{a}{2}\pm\frac{\sqrt{b}}{4}$.

The equation (\ref{lambda_equation}) has irregular singular point
$\lambda=\infty$ and regular singular point $\lambda=0$. One can introduce
``canonical" solutions $\Psi_k$ near infinity and $\Psi^0$ near the
point zero. The canonical asymptotics near infinity are
$$
\Psi_k(\lambda)=\Bigl(I+{\cal O}\bigl(\lambda^{-1}\bigr)\Bigr)
\exp\Bigl\{\Bigl(\frac{1}{8}\lambda^4+\frac{1}{2}\,x\lambda^2+
(\alpha-\beta)\ln\lambda\Bigr)\sigma_3\Bigr\},\quad
k\in{\Bbb Z}\,,$$
$$
\lambda\to\infty,\quad
\lambda\in\omega_k=\Bigl\{\lambda\in{\Bbb C}\,: \quad
\arg\lambda\in\bigl(-\frac{3\pi}{8}+\frac{\pi}{4}k;\
\frac{\pi}{8}+\frac{\pi}{4}k\bigr)\Bigr\},$$
moreover
$$
\Psi_{k+1}(\lambda)=\Psi_k(\lambda)S_k\, ,\quad
S_{2k-1}=\pmatrix{1&s_{2k-1}\cr 0&1\cr},\quad
S_{2k}=\pmatrix{1&0\cr s_{2k}&1\cr}.$$
The canonical solution near the point zero is described by
\begin{equation}\label{psi^0}
\Psi^0(\lambda)=\hat\Psi(\lambda)e^{\alpha\ln\lambda\sigma_3}P(\lambda),
\end{equation}
where
$$
P(\lambda)=\Bigl(I+j_{\pm}\ln\lambda\cdot\sigma_{\pm}\Bigr)
\exp\Bigl(\int^xuv\,dx\cdot\sigma_3\Bigr),$$
with the parameters $j_{\sigma}$, $\sigma\in\{+;-\}$, satisfying the
triviality conditions
$$
j_{\sigma}=0\quad\hbox{ if }\quad
{1\over 2}+\sigma\alpha\not\in{\Bbb N}\,,\quad\hbox{ i.e. }\quad
\alpha\notin\bigl\{\sigma\frac{1}{2};\sigma\frac{3}{2};\dots\bigr\},\quad
\sigma\in\{+;-\},$$
and $\hat\Psi(\lambda)$ holomorphic and invertible near the point
zero. In fact, the parameters $j_{\pm}$ can be expressed via the
coefficients of the system (\ref{lambda_equation}). For example,
\begin{eqnarray}\label{jp}
\hbox{ if }\quad
&&\alpha=\frac{1}{2}\,,\quad\hbox{ then }\quad
j_+=i(2xu+u'),
\\[4pt]\label{jm}
\hbox{and if}\quad
&&\alpha=-\frac{1}{2}\,,\quad\hbox{ then }\quad
j_-=i(2xv-v').
\end{eqnarray}

Generally spoken, the solution $\Psi^0(\lambda )$ is not fixed by the
asymptotics (\ref{psi^0}). Indeed,
\begin{description}
\item{1)} the asymptotics (\ref{psi^0}) is defined up to a right multiplier of
the form $C^{\sigma_3}$ with the arbitrary constant $C$;
\item{2)} if $2\alpha\in{\Bbb Z}\,$, one may add one of the columns of
(\ref{psi^0}) multiplied by an arbitrary coefficient
to another with preserving the asymptotics. It is equivalent to
multiplying of (\ref{psi^0}) from the right by a triangular matrix with the
unit diagonal. In the cases $\alpha\in{\Bbb Z}\,$ or
$\frac{1}{2}\pm\alpha\in{\Bbb N}\,$ with $j_{\pm}\neq 0$, the arbitrariness can
be eliminated by the additional condition
\begin{equation}\label{psi^0_jump_matrix}
\sigma_3\Psi^0(e^{i\pi}\lambda )\sigma_3=\Psi^0(\lambda )M
\end{equation}
with the jump matrix
\begin{equation}\label{M}
M=\Bigl(I+i\pi j_{\pm}e^{\mp2\int^xy(s)\,ds}\sigma_{\pm}\Bigr)
e^{i\pi\alpha\sigma_3}.
\end{equation}
If $\alpha -\frac{1}{2}\in{\Bbb Z}\,$ and $j_{\pm}=0$, the arbitrariness
preserves.
\end{description}

There is also the symmetry property
\begin{equation}\label{psi_k+4}
\sigma_3\Psi_{k+4}(e^{i\pi }\lambda )\sigma_3=
\Psi_k(\lambda)e^{i\pi(\alpha-\beta)\sigma_3},
\end{equation}
which yields
\begin{equation}\label{s_k+4}
S_{k+4}=e^{-i\pi (\alpha -\beta )\sigma _3}
\sigma _3S_k\sigma _3e^{i\pi (\alpha -\beta )\sigma _3}\, ,\quad
s_{k+4}=-s_{k}e^{(-1)^k2\pi i(\alpha -\beta )},
\end{equation}
and, with (\ref{psi^0_jump_matrix}) together, implies the so-called
semi-cyclic relation
\begin{equation}\label{cyclic}
ES_1S_2S_3S_4=\sigma _3M^{-1}E
e^{i\pi (\alpha -\beta )\sigma _3}\sigma _3
\end{equation}
where $E$ is the connection matrix
\begin{equation}\label{E}
E=\Psi ^0(\lambda )^{-1}\Psi _1(\lambda )=
\pmatrix{p&q\cr r&s\cr},\quad \det E=1\,.
\end{equation}
The connection matrix is as determined by (\ref{E}) as the solution $\Psi^0$
is determined, i.e.\ the matrix is fixed up to the left multiplier
$C^{\sigma_3}$ with the arbitrary constant $C$ and, if
$\frac{1}{2}\pm\alpha\in{\Bbb N}\,$, $j_{\pm}=0$, the matrix may be multiplied
by the arbitrary left upper (lower) triangular matrix.

The relation (\ref{cyclic}) can be considered as the system of four linear
homogeneous equations for entries $p$, $q$, $r$, $s$, $ps-qr=1$, of the
connection matrix $E$. The condition of its solvability (i.e.\ triviality of
the corresponding determinant) is equivalent to the equation
\begin{equation}\label{monodromy_surface}
\Bigl((1+s_1s_2)(1+s_3s_4)+s_1s_4\Bigr)e^{-i\pi(\alpha-\beta)}-
(1+s_2s_3)e^{i\pi(\alpha-\beta)}
=-2i\sin\pi\alpha\,,
\end{equation}
which is called the monodromy surface, so that only three of the Stokes
multipliers are independent from each other. The surface
(\ref{monodromy_surface}) has some special 1-dimensional submanifolds
defined by the equations
\begin{eqnarray}\label{special_points}
&&\alpha ={1\over 2}+n,\quad n\in {\Bbb Z}\,,\nonumber
\\[4pt]
&&s_1=-s_3e^{-i\pi\beta},\quad
s_4=-s_2e^{-i\pi\beta},\quad
1+s_2s_3=e^{i\pi\beta}.
\end{eqnarray}
It can be shown, for the non-special points of the surface
(\ref{monodromy_surface}), the connection matrix $E$ does not contain any
essential free parameter, but for the special points (\ref{special_points}),
it does, and this additional free parameter is the ratio of row components of
this matrix $E$. Thus the manifold of the monodromy data for
(\ref{lambda_equation}) is the surface (\ref{monodromy_surface}) with
${\Bbb CP}^1$ pasted to each special point (\ref{special_points}).

Note that the monodromy data including the Stokes
matrices $S_k$ or the Stokes multipliers $s_k$ are functions
of the coefficients of (\ref{lambda_equation}):
$$
S_k=S_k(x,u,v,\alpha ),\quad
s_k=s_k(x,u,v,\alpha ),\quad k\in {\Bbb Z}\,.$$
These functions possess the following symmetries:
\begin{equation}\label{s_symmetry_-u-v}
s_k(x,-u,-v,\alpha )=-s_k(x,u,v,\alpha ),
\quad k\in {\Bbb Z}\,;
\end{equation}
\begin{equation}\label{s_symmetry_bar}
\bar s_{-k}(\bar x,\bar u,\bar v,\bar \alpha )=
s_k(x,u,v,\alpha ),\quad k\in {\Bbb Z}\,,
\end{equation}
where the bar means the complex conjugation;
the gauge symmetry
\begin{equation}\label{s_symmetry_gauge}
S_k(x,e^au,e^{-a}v,\alpha )=e^{{a\over 2}\sigma _3}
S_k(x,u,v,\alpha )e^{-{a\over 2}\sigma _3},\quad a\in {\Bbb C}\,,
\quad k\in {\Bbb Z}\,;
\end{equation}
and the rotation symmetry
\begin{eqnarray}\label{s_symmetry_rotation}
&&S_{k+n}\bigl(e^{i\frac{\pi}{2}n}x,\tau ^n(e^{i\frac{\pi}{4}n}u,
e^{i\frac{\pi}{4}n}v),e^{i\pi n}\alpha\bigr)=
\\[4pt]
&&=(\sigma _2)^ne^{-\frac{i\pi}{4}n(\alpha-\beta)\sigma_3}
S_k(x,u,v,\alpha)
e^{\frac{i\pi}{4}n(\alpha-\beta)\sigma_3}(\sigma_2)^n,
\quad k,n\in {\Bbb Z}\,,\nonumber
\end{eqnarray}
where $\tau $ is permutation $\tau (u,v)=(v,u)$. The symmetry
(\ref{s_symmetry_rotation}) can be treated as a B\"acklund transformation in
the following sense: if some function $y(x)=f(x,\alpha,\beta,\{s_k\})$
describes the Painlev\'e function then the function
\begin{equation}\label{y_symmetry_rotation}
\tilde y(\tilde x)=e^{i\frac{\pi}{2}n}
f\Bigl(e^{-i\frac{\pi}{2}n}\tilde x,(-1)^n\tilde\alpha,(-1)^n\tilde\beta,
\{(-1)^ne^{i\frac{\pi}{2}(-1)^kn(\alpha-\beta)} s_{k+n}\}\Bigr)
\end{equation}
describes another solution of P4 of the new variable
$\tilde x=e^{i\frac{\pi}{2}n}x$ and the new parameters
$\tilde\alpha=(-1)^n\alpha$, $\tilde\beta=(-1)^n\beta$. The symmetry
(\ref{s_symmetry_bar}) may be treated in the similar way.

The linear system (\ref{lambda_equation}) maps a set of the coefficients
$x$, $u$, $v$, $u'$, $v'$ onto a manifold of the monodromy data described
above. So, the general (not special) points of the 3-d complex monodromy
surface (\ref{monodromy_surface}) are in one-to-one correspondence with the
3-d complex set of the functions $u$, $v$, $u'$, $v'$ depending on $x$
in accord with (\ref{system_uv}), (\ref{beta}). At the special points
(\ref{special_points}), the points of locally 2-d complex monodromy
manifold (one of the Stokes multipliers and a ratio of the connection matrix
entries) parameterize 2-d complex set in the space of the functions $u$,
$v$, $u'$, $v'$ with the restriction (\ref{beta}) and equation $j_{\pm}=0$
together. (Note the proposed in \cite{MCB} parametrization of 2-d set of the
Painlev\'e functions $y$, $y'$ by the points of 4-d monodromy manifold
is less convenient.)

The equation (\ref{s_symmetry_gauge}) means the fourth Painlev\'e transcendent
$y=uv$ is invariant about the gauge transformation in contrast to the
coefficients $u$, $v$, so that any solution of P4 corresponds to an orbit of
the 1-parametric group of the gauge transformations of the monodromy data
manifold.

\section{B\"acklund and Schlesinger transformations.}
\label{backlund_schlesinger}

In what follows, the crucial role is played by the group of the B\"acklund
transformations of P4 (\ref{p4}) considered in refs.\ \cite{L, FMA, M, FA, Gr}
generated by the group of Schlesinger transformations (see ref.\
\cite{JM}) of the linear system (\ref{lambda_equation}). The elementary
Schlesinger transformation of the function $\Psi$ preserving the monodromy
data except the exponents of the formal monodromy $\alpha$ near the point
zero or $\alpha-\beta$ near infinity is defined as $\tilde\Psi=R\Psi$ with
the rational matrix function $R$ of one of the following forms:
\begin{eqnarray}\label{schlesinger}
&&R_0^+=I+\frac{i}{\lambda}\cdot\frac{1+2\alpha}{2xv-v'}\,\sigma_+\,,\nonumber
\quad
R_0^-=I+\frac{i}{\lambda}\cdot\frac{1-2\alpha}{2xu+u'}\,\sigma_-\, ,
\\[4pt]
&&R_{\infty}^+=\pmatrix{\lambda& iu\cr \frac{i}{u}&0\cr},
\quad
R_{\infty}^-=\pmatrix{0&\frac{1}{iv}\cr -iv&\lambda\cr}.
\end{eqnarray}

The transformed connection coefficient $\tilde{\cal A}$ is described by
(\ref{lambda_equation}) with coefficients $u$, $u'$, $v$, $v'$, $\alpha$
replaced by $\tilde u$, $\tilde u'$, $\tilde v$, $\tilde v'$,
$\tilde\alpha$:
\begin{eqnarray}\label{backlund_0^+}
\hbox{if }\quad
&&R=R_0^+\colon\nonumber
\\[4pt]
&&\tilde\alpha=-\alpha-1,\quad
\tilde\alpha-\tilde\beta=\alpha-\beta,\nonumber
\\[4pt]
&&\tilde u=u-\frac{1+2\alpha}{2xv-v'},\quad
\tilde v=v,\nonumber
\\[4pt]
&&\tilde u'=u'-\frac{(1+2\alpha)v}{2xv-v'}(u+\tilde u),\quad
\tilde v'=v',\nonumber
\\[4pt]
\hbox{so that }\quad
&&\tilde y=
R_0^+[y]\equiv
y+\frac{2(1+2\alpha)y}{y'-y^2-2xy-\beta}\, ,
\\[4pt]\label{backlund_0^-}
\hbox{if }\quad
&&R=R_0^-\colon\nonumber
\\[4pt]
&&\tilde\alpha=-\alpha+1,\quad
\tilde\alpha-\tilde\beta=\alpha-\beta,\nonumber
\\[4pt]
&&\tilde u=u,\quad
\tilde v=v+\frac{1-2\alpha}{2xu+u'},\nonumber
\\[4pt]
&&\tilde u'=u',\quad
\tilde v'=v'-\frac{(1-2\alpha)u}{2xu+u'}(v+\tilde v),\nonumber
\\[4pt]
\hbox{so that }\quad
&&\tilde y=
R_0^-[y]\equiv y+\frac{2(1-2\alpha)y}{y'+y^2+2xy+\beta}\, ,
\\[4pt]\label{backlund_infty^+}
\hbox{if }\quad
&&R=R_{\infty}^+\colon\nonumber
\\[4pt]
&&\tilde\alpha=-\alpha,\quad
\tilde\alpha-\tilde\beta=\alpha-\beta+1,\nonumber
\\[4pt]
&&\tilde u=u'-u^2v,\quad
\tilde v=\frac{1}{u},\nonumber
\\[4pt]
&&\tilde u'=-2xu'-u^2v'+4xu^2v-(1+2\alpha)u,\quad
\tilde v'=-\frac{u'}{u^2},\nonumber
\\[4pt]
\hbox{so that }\quad
&&\tilde y=
R_{\infty}^+[y]\equiv\frac{y'}{2y}-\frac{1}{2}y-x+\frac{\beta}{2y}\, ,
\\[4pt]\label{backlund_infty^-}
\hbox{if }\quad
&&R=R_{\infty}^-\colon\nonumber
\\[4pt]
&&\tilde\alpha=-\alpha,\quad
\tilde\alpha-\tilde\beta=\alpha-\beta-1,\nonumber
\\[4pt]
&&\tilde u=\frac{1}{v},\quad
\tilde v=-v'-uv^2,\nonumber
\\[4pt]
&&\tilde u'=-\frac{v'}{v^2},\quad
\tilde v'=-2xv'-u'v^2-4xuv^2-(1-2\alpha)v,\nonumber
\\[4pt]
\hbox{so that }\quad
&&\tilde y=
R_{\infty}^-[y]\equiv-\frac{y'}{2y}-\frac{1}{2}y-x+\frac{\beta}{2y}\, ,
\end{eqnarray}
The relations (\ref{backlund_0^+})--(\ref{backlund_infty^-}) are called
the B\"acklund transformations of the fourth Painlev\'e transcendent.

The B\"acklund transformations (\ref{backlund_infty^+}),
(\ref{backlund_infty^-}) are found in ref.\ \cite{L}. Two transformations
(\ref{backlund_0^+}), (\ref{backlund_0^-}) with the corresponding Schlesinger
transformations are obtained in ref.\ \cite{Kit(1988)}. Other B\"acklund
transformations found in refs.\ \cite{A, FMA, M} can be described as some
superpositions of these basic transformations.

\section{Classical solutions}\label{classical_solutions}

It is well known \cite{L} that P4 has two series of rational 0-solutions
existing for integer and only integer value of the parameter
$a=2\alpha-\frac{\beta}{2}$:

1) $-\frac{2}{3}\,x+\frac{P_{n-1}(x)}{Q_n(x)}$ generated by the actions of
the B\"acklund transformations on the ``seed" solution of (\ref{p4})
$y=-\frac{2}{3}\,x$ for $a=0$, $b=\frac{4}{9}$. Each of the rational solutions
corresponds to the Stokes multipliers satisfying the conditions
\begin{equation}\label{sk_rat_2/3}
1+s_ks_{k+1}=0,\quad k\in{\Bbb Z}\, .
\end{equation}
These rational solutions are some limiting cases of 2-parametric
transcendent Painlev\'e functions \cite{Kap}.

2) Other family of rational solutions described in ref.\ \cite{L, GL, M} is
a collection of two forms: a) $\frac{P_{n-1}(x)}{Q_n(x)}\,$, and
b) $-2x+\frac{P_{n-1}(x)}{Q_n(x)}\,$. In accord with ref.\ \cite{L}, all these
solutions can be generated from the solution $y_0=-2x$, existing for $a=0$,
$b=4$ (i.e., $\beta=\pm2$, $\alpha=\pm\frac{1}{2}$). The ``seed" solution
$y=-2x$ corresponds to the Stokes multipliers
\begin{equation}\label{sk_rat20}
s_{2k-1}=0,\quad s_{2k}+s_{2k+2}=0,\quad
k\in{\Bbb Z}\, ,
\quad\hbox{if}\quad\alpha=\frac{1}{2}\,,\quad\beta=2,
\end{equation}
or
\begin{equation}\label{sk_rat2}
s_{2k}=0,\quad s_{2k-1}+s_{2k+1}=0,\quad
k\in{\Bbb Z}\, ,
\quad\hbox{if}\quad\alpha=-\frac{1}{2}\,,\quad\beta=-2.
\end{equation}
Its B\"acklund transformations are characterized by the same Stokes
multiplier values but some other $\alpha$ and $\beta$. Note also, all the
rational functions are some limiting cases of the classical solutions of P4
(see below). The corresponding $\Psi$ function can be expressed via the
Weber-Hermite functions.

The simplest of the so-called classical solutions of P4 satisfy the Riccati
equation \cite{A, GL} $y'=\sigma\bigl(y^2+2xy\bigr)+q$, where $\sigma^2=1$,
so the parameters of (\ref{p4}) are $a=-\sigma\bigl(\frac{q}{2}+1\bigr)$,
$b=q^2$ (so, for $q=-2$, $a=0$, there are two different kinds of the simplest
classical solutions corresponding to both values of $\sigma=\pm1$; for
details, see also \cite{UW}). In terms
of the monodromy data, the classical solutions correspond
to the special linear submanifolds of the monodromy manifold. In fact, there
are two kinds of such submanifolds (cf.\ \cite{MCB}). The first kind of the
submanifolds encloses all the projective complex spaces ${\Bbb CP}^1$ pasted
to each of the special points (\ref{special_points}) of the monodromy surface
(\ref{monodromy_surface}) and in terms of the parameters of the
$\lambda$-equation corresponding to the trivial values $j_{\pm}=0$. The second
kind of the submanifolds is distinguished by the system of equations
$s_{2k-1}=0$ or $s_{2k}=0$.

\smallskip
1) Let us consider the special points corresponding to $\alpha=\frac{1}{2}\,$.
As it is said above, these special points are characterized by the equation
$j_+=0$, or $u'+2xu=0$ (\ref{jp}), that yields $u=Ce^{-x^2}$ with an arbitrary
constant $C$. Now, the definition of $\beta$ (\ref{beta}) gives the Riccati
equation for the function $v$: $v'+uv^2+\frac{\beta}{u}=0$. Substitution
$v=\frac{1}{u}\,\frac{z'}{z}$ produces the linear equation
\begin{equation}\label{z_p}
z''+2xz'+\beta z=0\,.
\end{equation}
The Painlev\'e function (\ref{y}) $y=uv=z'/z$ solves the Riccati equation
\begin{equation}\label{ric_mm}
y'=-y^2-2xy-\beta\,.
\end{equation}
As to the solution $\Psi(\lambda)$ of the equation (\ref{lambda_equation}),
it can be expressed in terms of the Weber-Hermite functions.

\smallskip
2) For the opposite parameter value $\alpha=-\frac{1}{2}$, the special
points are characterized by the equation $j_-=0$, or $v'-2xv=0$ (\ref{jm}),
so that $v=Ce^{x^2}$ with an arbitrary constant $C$. Another coefficient $u$
satisfies the Riccati equation $u'-u^2v-\frac{\beta}{v}=0$ which is
linearized by the change $u=-\frac{1}{v}\,\frac{z'}{z}$:
\begin{equation}\label{z_m}
z''-2xz'+\beta z=0\,.
\end{equation}
Now, $y=uv=-z'/z\,$ satisfies the Riccati equation
\begin{equation}\label{ric_pp}
y'=y^2+2xy+\beta\,.
\end{equation}
Similarly to the previous case, the equation (\ref{lambda_equation}) can be
solved explicitly in terms of the parabolic cylinder functions.

\smallskip
3) Next linearization of the monodromy manifold takes place under
lower triangular reduction of the $\lambda$-equation (\ref{lambda_equation})
obtained by means of the restriction $u\equiv0$. The definition (\ref{beta})
gives immediately $\beta=0$, while after (\ref{system_uv}), another
coefficient $v$ satisfies the linear equation
\begin{equation}\label{v_m}
v''-2xv'+\bigl(2\alpha-1\bigr)v=0\,.
\end{equation}
As easy to see, the triangular system (\ref{lambda_equation}) can be solved
in quadratures and yields the lower triangular $\Psi$ function and the lower
triangularity of all the Stokes matrices, the monodromy matrix and the
connection matrix, so that $s_{2k-1}=0$, $k\in{\Bbb Z}\,$, $j_+=0$, $p=s=1$,
$q=0$, moreover the equation (\ref{monodromy_surface}) holds identically while
(\ref{cyclic}) reduces to the only equation
$s_2+s_4=-r\bigl(1+e^{2\pi i\alpha}\bigr)$ if
$\frac{1}{2}-\alpha\notin{\Bbb N}\,$, or
$s_2+s_4=-i\pi j_-\exp\bigl(2\int^xy\,dx\bigr)$ if
$\frac{1}{2}-\alpha\in{\Bbb N}\,$.
By definition (\ref{y}), the case corresponds to the trivial Painlev\'e
function $y\equiv0$ which does not allow to apply the transformations
(\ref{backlund_0^+})--(\ref{backlund_infty^-}). However, the described
coefficients $u$, $v$ can be affected by the B\"acklund transformations
$R_0^+$ (\ref{backlund_0^+}) and $R_{\infty}^-$ (\ref{backlund_infty^-}).
Modified values are as follows:
$$
R_0^+\colon\,\qquad
\tilde\alpha=-\alpha-1\,,\quad
\tilde\beta=-2\alpha-1\,,$$
$$
\tilde u=-\frac{1+2\alpha}{2xv-v'}\,,\quad
\tilde v=v,\quad
\tilde y=\tilde u\tilde v=-\frac{1+2\alpha}{2xv-v'}\,v\,,$$
so that
\begin{equation}\label{ric_pm}
\tilde y'=\tilde y^2+2x\tilde y-\tilde\beta\,.
\end{equation}
Another B\"acklund transformation $R_{\infty}^-$ gives
$\tilde\alpha=-\alpha$,\quad $\tilde\beta=-2\alpha+1$,\quad
$\tilde u=\frac{1}{v}$,\quad $\tilde v=-v'$,\quad
$\tilde y=\tilde u\tilde v=-\frac{v'}{v}$, and $\tilde y$ satisfies the
Riccati equation (\ref{ric_pm}).

\smallskip
4) The upper-triangular reduction takes place if $v\equiv0$, so that
$\beta=0$, $y\equiv0$, $s_{2k}=0$, $k\in{\Bbb Z}\,$, $j_-=0$, $p=s=1$, $r=0$,
while
$s_1+s_3=-q\bigl(1+e^{-2\pi i\alpha}\bigr)$ if
$\frac{1}{2}+\alpha\notin{\Bbb N}\,$, or
$s_1+s_3=-i\pi j_+\exp\bigl(-2\int^xy\,dx\bigr)$ if
$\frac{1}{2}+\alpha\in{\Bbb N}\,$.
The coefficient function $u$ satisfies the linear equation
\begin{equation}\label{u_p}
u''+2xu'+\bigl(2\alpha+1\bigr)u=0\,.
\end{equation}
The coefficients affected by the B\"acklund transformations $R_0^-$
(\ref{backlund_0^-}) and $R_{\infty}^+$ (\ref{backlund_infty^+}) are as
follows:
$$
R_0^-\colon\,\qquad
\tilde\alpha=-\alpha+1\,,\quad
\tilde\beta=-2\alpha+1\,,$$
$$
\tilde u=u\,,\quad
\tilde v=\frac{1-2\alpha}{2xu+u'}\,,\quad
\tilde y=\tilde u\tilde v=\frac{1-2\alpha}{2xu+u'}\,u\,,$$
so that
\begin{equation}\label{ric_mp}
\tilde y'=-\tilde y^2-2x\tilde y+\tilde\beta\,.
\end{equation}
Another transformation $R_{\infty}^+$ yields
$\tilde\alpha=-\alpha$,\quad
$\tilde\beta=-2\alpha-1$,\quad
$\tilde u=u'$,\quad
$\tilde v=\frac{1}{u}$,\quad
$\tilde y=\tilde u\tilde v=\frac{u'}{u}$,
and $\tilde y$ satisfies the same Riccati equation (\ref{ric_mp}).

\smallskip
Note that the solutions of the equations (\ref{ric_mm}) and (\ref{ric_pp}) are
related with each other. Namely, if $y$ solves the Riccati
equation (\ref{ric_mm}) then its B\"acklund transformation $R_{\infty}^-$
reduced to $\tilde y=\beta/y$ solves the Riccati equation (\ref{ric_pp}).
Similarly, if $y=f(x,\beta)$ is a solution of (\ref{ric_mm}) then
$\tilde y=\pm if(\pm ix,-\beta)$ solves (\ref{ric_pp}). The solutions of the
Riccati equations (\ref{ric_pm}) and (\ref{ric_mp}) can be obtained via
simple substitution in solutions of (\ref{ric_pp}) and (\ref{ric_mm}) the
parameter $-\tilde\beta$ instead of $\beta$.

Let us describe the global asymptotics of the presented classical solutions.
Because of the said above we can restrict ourselves in
the case (\ref{ric_mm}), when the substitution $y=z'/z$ yield
the linear equation (\ref{z_p}). As easy to see \cite{GL}, the change
\begin{equation}\label{z_to_wm}
z=e^{-\frac{x^2}{2}}w
\end{equation}
transform the equation (\ref{z_p}) into the parabolic cylinder equation
\begin{equation}\label{parabolic_cylinder_m}
w''+\bigl(\beta-1-x^2\bigr)w=0\,.
\end{equation}
Using the results on the asymptotic properties of the Weber-Hermite
functions presented in the reference book \cite{BE}, we can introduce the
vectors of fundamental solutions of the equation (\ref{parabolic_cylinder_m})
$W_k=\bigl(w_1^{(k)};w_2^{(k)}\bigr)$
\begin{eqnarray}\label{asymptotics_w}
&&W_k=\Bigl(e^{\frac{x^2}{2}-\frac{\beta}{2}\ln x}
\bigl(1+{\cal O}(x^{-2})\bigr)\,;\
e^{-\frac{x^2}{2}+(\frac{\beta}{2}-1)\ln x}
\bigl(1+{\cal O}(x^{-2})\bigr)\Bigr),\nonumber
\\[4pt]
&&\arg x\in\Bigl(\frac{\pi}{2}(k-1)\,;\
\frac{\pi}{2}k\Bigr),\quad
|x|\to\infty\,.
\end{eqnarray}
The canonical vectors are related with each other by the Stokes matrices
\begin{eqnarray}\label{gk}
&&W_{k+1}(x)=W_k(x)G_k\,,
\\[4pt]
&&G_{2k}=\pmatrix{1&0\cr g_{2k}&1\cr},\quad
G_{2k-1}=\pmatrix{1&g_{2k-1}\cr 0&1\cr},\nonumber
\\[4pt]
&&g_0=-i\sqrt\pi\,\frac{2^{\frac{\beta}{2}}}
{\Gamma\bigl(\frac{\beta}{2}\bigr)}\,,\quad
g_1=\sqrt\pi\,\frac{2^{1-\frac{\beta}{2}}\,e^{-i\pi(1-\frac{\beta}{2})}}
{\Gamma\bigl(1-\frac{\beta}{2}\bigr)},\nonumber
\\[4pt]
&&g_{2k+2}=-g_{2k}e^{-i\pi\beta}\,,\quad
g_{2k+1}=-g_{2k-1}e^{i\pi\beta}\,.\nonumber
\end{eqnarray}
In accord with (\ref{z_to_wm}), the corresponding fundamental vector of the
solutions of (\ref{z_p}), i.e.\ $Z_k=e^{-x^2/2}W_k$ yield the ``generating"
matrix for the function $y=z'/z$:
\begin{equation}\label{asymptotics_z}
Z_k=\pmatrix{z_1'&z_2'\cr z_1&z_2\cr}=
\pmatrix{-\frac{\beta}{2}\,x^{-\frac{\beta}{2}-1}\bigl(1+{\cal O}(x^{-2})&
-2e^{-x^2}x^{\frac{\beta}{2}}\bigl(1+{\cal O}(x^{-2})\bigr)\cr
x^{-\frac{\beta}{2}}\bigl(1+{\cal O}(x^{-2})\bigr)&
e^{-x^2}x^{\frac{\beta}{2}-1}\bigl(1+{\cal O}(x^{-2})\bigr)\cr}.\nonumber
\end{equation}
The classical solution of P4 satisfying the Riccati equation (\ref{ric_mm})
$y'=-y^2-2xy-\beta$,
\begin{equation}\label{yz'z}
y=\frac{\nu_1z_1'+\nu_2z_2'}{\nu_1z_1+\nu_2z_2}
\end{equation}
goes from (\ref{asymptotics_z}) after multiplication in the column of
constants $\nu_{1,2}$ and dividing the first entry to the second. As easy to
see, this exact solution behaves asymptotically as $|x|\to\infty$,
$\arg x\neq\frac{\pi}{4}+\frac{\pi}{2}n$, $n\in{\Bbb Z}\,$, like
\begin{equation}\label{mains}
y\sim-2x\quad\hbox{ or }\quad
y\sim-\frac{\beta}{2x}
\end{equation}
In fact, these asymptotics are described by means of the asymptotic
power series for $z_i'/z_i$ and an exponentially small perturbation depending
on the arbitrary parameter $\nu_1/\nu_2$. (Along the excluded rays
$\arg x=\frac{\pi}{4}+\frac{\pi}{2}n$, the ratio (\ref{yz'z}) is described
asymptotically by some trigonometric tangent.) In more details, if
$\alpha=\frac{1}{2}$, the classical solution of P4 (\ref{p4}) corresponding
to the special point (\ref{special_points}) satisfies the Riccati equation
(\ref{ric_mm}) $y'+y^2+2xy+\beta=0$ and behaves asymptotically as follows:
\begin{description}
\item{i)} $\arg x\in\bigl(-\frac{\pi}{4}\,;\ \frac{\pi}{4}\bigr),\quad
\nu_1\neq0$:
\begin{equation}\label{cy1}
y=-\frac{\beta}{2x}\Bigl(1+{\cal O}\bigl(x^{-2}\bigr)\Bigr)-
2\Bigl(\frac{\nu_2}{\nu_1}-\Theta(\arg x)g_0\Bigr)x^{\beta}e^{-x^2}
\Bigl(1+{\cal O}\bigl(x^{-2}\bigr)\Bigr)\,;
\end{equation}
\item{ii)} $\arg x\in\bigl(\frac{\pi}{4}\,;\ \frac{3\pi}{4}\bigr),\quad
\nu_2-\nu_1g_0\neq0$:
\begin{equation}\label{cy2}
y=-2x\Bigl(1+{\cal O}\bigl(x^{-2}\bigr)\Bigr)+
2\Bigl(\frac{\nu_1}{\nu_2-\nu_1g_0}-\Theta\bigl(\arg x-\frac{\pi}{2}\bigr)g_1\Bigr)
x^{2-\beta}e^{x^2}
\Bigl(1+{\cal O}\bigl(x^{-2}\bigr)\Bigr)\,;
\end{equation}
\item{iii)} $\arg x\in\bigl(\frac{3\pi}{4};\ \frac{5\pi}{4}\bigr),\quad
\nu_1e^{i\pi\beta}-\nu_2g_1\neq0$:
\begin{equation}\label{cy3}
y=-\frac{\beta}{2x}\Bigl(1+{\cal O}\bigl(x^{-2}\bigr)\Bigr)-
2\Bigl(\frac{\nu_2-\nu_1g_0}{\nu_1e^{i\pi\beta}-\nu_2g_1}-\Theta(\arg x-\pi)g_2\Bigr)
x^{\beta}e^{-x^2}
\Bigl(1+{\cal O}\bigl(x^{-2}\bigr)\Bigr)\,;
\end{equation}
\item{iv)} $\arg x\in\bigl(\frac{5\pi}{4}\,;\ \frac{7\pi}{4}\bigr),\quad
\nu_2\neq0$:
\begin{equation}\label{cy4}
y=-2x\Bigl(1+{\cal O}\bigl(x^{-2}\bigr)\Bigr)+
2\Bigl(\frac{\nu_1e^{i\pi\beta}-\nu_2g_1}{\nu_2e^{-i\pi\beta}}-
\Theta\bigl(\arg x-\frac{3\pi}{2}\bigr)g_3\Bigr)x^{2-\beta}e^{x^2}
\Bigl(1+{\cal O}\bigl(x^{-2}\bigr)\Bigr)\,,
\end{equation}
where
$$
\Theta(z)=\cases{0\quad\hbox{ if }\quad z<0, \cr
                {1\over 2}\quad \hbox{ if }\quad z=0, \cr
                 1\quad\hbox{ if }\quad z>0. \cr }$$
\end{description}
Excluded values of the parameters correspond to the opposite algebraic terms
(see (\ref{mains})) and absence of any exponentially small perturbation.
For example, if $\nu_1=0$, then
$y=-2x\Bigl(1+{\cal O}\bigl(x^{-2}\bigr)\Bigr)$ for
$\arg x\in\bigl(-\frac{\pi}{4}\,;\ \frac{\pi}{4}\bigr)$.

On the rays $\arg x=\frac{\pi}{2}\,k$, $k\in{\Bbb Z}\,$, the asymptotic formulae
(\ref{cy1})--(\ref{cy4}) demonstrate the quasi-linear Stokes phenomenon
imposed by the Stokes property of the Weber-Hermite functions which consists
in a jump of the exponentially small term while the independent variable
crosses the Stokes ray (see, e.g. \cite{B}).

\section{Degenerated asymptotic solutions\hfill\break
and the B\"acklund transformations.}\label{degenerated_solutions}

The results of the previous section on the asymptotics of the classical
solutions give us a clue to look for the formal asymptotic solutions of
the fourth Painlev\'e equation (\ref{p4}) in the form
\begin{equation}\label{formal_form}
\sum_{k=0}^{\infty}e^{\vartheta_k(x)}\sum_{n=0}^{\infty}a_{nk}x^{-n}\, ,
\end{equation}
with $\vartheta_k(x)=-kx^2+b_k\ln x$ and some constants $a_{nk}$, $b_k$.
The terms of the formal series can be evaluated recursively by use of any
automatic system of analytic calculations. Several first terms of the formal
asymptotic expansions below are calculated by use of MATHEMATICA~2.2 of
Wolfram Research, Inc. (for what follows, it is enough to keep two of the
successive exponentially small terms).

{\it i)} 0-parameter asymptotics of a background for 2-parameter oscillating
asymptotic solution:
\begin{equation}\label{y_2/3}
y_{2/3}(x,a,b)=-\frac{2x}{3}+\frac{a}{x}+\frac{-4-12a^2+9b}{16x^3}+
{\cal O}\bigl(\frac{1}{x^5}\bigr);
\end{equation}

\smallskip
{\it ii)} 1-parameter asymptotics approaching $-2x$:
\begin{eqnarray}\label{y_2}
&&y_2(x,a,b,c)=-2x-\frac{a}{x}+\frac{4 + 12a^2 - b}{16x^3}+
\frac{-44a - 36a^3 + 5ab}{32x^5}+{\cal O}\bigl(\frac{1}{x^7}\bigr)+\nonumber
\\[4pt]
&&+cx^{-2a}e^{-x^2}\Bigl\{1+\frac{-12-8a-28a^2+3b}{32x^2}+
{\cal O}\bigl(\frac{1}{x^4}\bigr)\Bigr\}-\nonumber
\\[4pt]
&&-\frac{c^2}{2}x^{-1-4a}e^{-2x^2}\Bigl\{1+\frac{-20-24a-28a^2+3b}{16x^2}+
{\cal O}\bigl(\frac{1}{x^4}\bigr)\Bigr\};
\end{eqnarray}

\smallskip
{\it iii)} 1-parameter decreasing asymptotics:
\begin{eqnarray}\label{y_pm}
&&y_{\pm}(x,a,\sqrt b,c)=\pm\frac{\sqrt b}{2x}\pm\frac{(a\mp\sqrt b)\sqrt b}
{4x^3}\pm\nonumber
\\[4pt]
&&\pm\frac{\sqrt b\,(12+12a^2\mp32a\sqrt b+17b)}{64x^5}+
{\cal O}\bigl(\frac{1}{x^7}\bigr)+\nonumber
\\[4pt]
&&+cx^{-1+a\mp\frac{3\sqrt b}{2}}e^{-x^2}
\Bigl\{1+\nonumber
\\[4pt]
&&+\frac{1}{32x^2}\bigl(-12+16a-4a^2\mp20\sqrt b\pm
24a\sqrt b-15b\bigr)+
{\cal O}\bigl(\frac{1}{x^4}\bigr)
\Bigr\}+\nonumber
\\[4pt]
&&+\frac{c^2}{2}x^{-3+2a\mp3\sqrt b}e^{-2x^2}
\Bigl\{1+\nonumber
\\[4pt]
&&+\frac{-36+32a-4a^2\mp36\sqrt b\pm24a\sqrt b-15b}{16x^2}
+{\cal O}\bigl(\frac{1}{x^4}\bigr)\Bigr\}.
\end{eqnarray}

\smallskip
The formal solutions $y_+(x,a,\sqrt b,c)=y_-(x,a,-\sqrt b,c)$ coincide with
each other if $b=0$. This asymptotic solution is denoted as $y_0$ and is given
by
\begin{eqnarray}\label{y_0}
&&y_0(x,a,c)=cx^{-1+a}e^{-x^2}
\Bigl\{1+\frac{-3+4a-a^2}{8x^2}+
{\cal O}\bigl(\frac{1}{x^4}\bigr)
\Bigr\}+\nonumber
\\[4pt]
&&+\frac{c^2}{2}x^{-3+2a}e^{-2x^2}
\Bigl\{1+\frac{-9+8a-a^2}{4x^2}
+{\cal O}\bigl(\frac{1}{x^4}\bigr)\Bigr\}.
\end{eqnarray}

It is readily seen that the B\"acklund transformation group generators
(\ref{backlund_0^+})--(\ref{backlund_infty^-}) preserve the set of the formal
asymptotic solutions (\ref{y_2/3})--(\ref{y_0}). In particular,
$$
\tilde y_{2/3}(x,a,b)\equiv R[y_{2/3}(x,a,b)]=y_{2/3}(x,\tilde a,\tilde b)$$
for any B\"acklund transformation $R$. The general assertion is also pure
algebraic and can be checked directly. Listed below actions of the
B\"acklund transformations (\ref{backlund_0^+})--(\ref{backlund_infty^-})
are obtained by use of MATHEMATICA~2.2. Here, $b=\beta^2$,
$a=2\alpha-\frac{\beta}{2}$, $\tilde y=R[y]$:
\begin{eqnarray}\label{R+8_action}
R^+_{\infty}\colon
&&\tilde\alpha=-\alpha\,,\quad
\tilde\beta=-2\alpha+\beta-1\, ,\nonumber
\\[4pt]
&&\tilde y_2(x,a,b,c)=y_-\bigl(x,\tilde a,\tilde\beta,
\frac{c}{4}(2\alpha-\beta+1)\bigr)\, ,\nonumber
\\[4pt]
&&\tilde y_-(x,a,\beta,c)=y_2(x,\tilde a,\tilde b,-c),\nonumber
\\[4pt]
&&\tilde y_+(x,a,\beta,c)=y_+\bigl(x,\tilde a,\tilde\beta,
-\frac{4c}{\beta}\bigr),\nonumber
\\[4pt]
&&\tilde y_0(x,a,c)=y_2(x,\tilde a,\tilde b,-c);
\\[8pt]\label{R-8_action}
R^-_{\infty}\colon
&&\tilde\alpha=-\alpha\, ,\quad
\tilde\beta=-2\alpha+\beta+1\, ,\nonumber
\\[4pt]
&&\tilde y_2(x,a,b,c)=y_-(x,\tilde a,\tilde\beta,-c)\, ,\nonumber
\\[4pt]
&&\tilde y_-(x,a,\beta,c)=y_2\bigl(x,\tilde a,\tilde b,
-\frac{4c}{\beta}\bigr),\nonumber
\\[4pt]
&&\tilde y_+(x,a,\beta,c)=y_+\bigl(x,\tilde a,\tilde\beta,
\frac{c}{4}(2\alpha-\beta-1)\bigr),\nonumber
\\[4pt]
&&\tilde y_0(x,a,c)=y_+\bigl(x,\tilde a,\tilde\beta,
\frac{c}{4}(2\alpha-1)\bigr);
\\[8pt]\label{R+0_action}
R^+_0\colon
&&\tilde\alpha=-\alpha-1\, ,\quad
\tilde\beta=-2\alpha+\beta-1\, ,\nonumber
\\[4pt]
&&\tilde y_2(x,a,b,c)=y_-\bigl(x,\tilde a,\tilde\beta,
-\frac{c}{16}(2\alpha-\beta+1)(2\alpha+1)\bigr)\, ,\nonumber
\\[4pt]
&&\tilde y_-(x,a,\beta,c)=y_2\Bigl(x,\tilde a,\tilde b,
-\frac{16c}{\beta(2\alpha+1)}\Bigr),\nonumber
\\[4pt]
&&\tilde y_+(x,a,\beta,c)=y_+\bigl(x,\tilde a,\tilde\beta,c\bigr),\nonumber
\\[4pt]
&&\tilde y_0(x,a,c)=y_+(x,\tilde a,\tilde\beta,c);
\\[8pt]\label{R-0_action}
R^-_0\colon
&&\tilde\alpha=-\alpha+1\, ,\quad
\tilde\beta=-2\alpha+\beta+1\, ,\nonumber
\\[4pt]
&&\tilde y_2(x,a,b,c)=y_-\bigl(x,\tilde a,\tilde\beta,
\frac{4c}{2\alpha-1}\bigr)\, ,\nonumber
\\[4pt]
&&\tilde y_-(x,a,\beta,c)=y_2\bigl(x,\tilde a,\tilde b,
-\frac{c}{4}(2\alpha-1)\bigr),\nonumber
\\[4pt]
&&\tilde y_+(x,a,\beta,c)=y_+\bigl(x,\tilde a,\tilde\beta,
-\frac{c}{\beta}(2\alpha-\beta-1)\bigr),\nonumber
\\[4pt]
&&\tilde y_0(x,a,c)=y_2\bigl(x,\tilde a,\tilde b,
-\frac{c}{4}(2\alpha-1)\bigr).
\end{eqnarray}

The invariance of the degenerated solution set allows us to introduce such
a dependence of the multipliers $c$ of the exponentially small terms on the
parameters $\alpha$ and $\beta$ to ensure their invariance about the
B\"acklund transformations action. Let us provide the original
parameters $c$ of the asymptotic solutions $y_t$, $t\in\{2;+;-;0\}$
(\ref{y_2})--(\ref{y_0}) by the same indices $t$. Then
\begin{eqnarray}\label{c_2}
&&c_2(\alpha,\beta)=
\frac{2^{-2\alpha+\frac{\beta}{2}}}
{\Gamma\bigl(\frac{1}{2}-\alpha\bigr)
\Gamma\bigl(\frac{1}{2}-\alpha+\frac{\beta}{2}\bigr)}\,
f_2\bigl(\alpha\bigr)g_2\bigl(\alpha-\frac{\beta}{2}\bigr)\,,
\\[4pt]\label{c_-}
&&c_-(\alpha,\beta)=\frac{2^{\alpha+\frac{\beta}{2}}}
{\Gamma\bigl(\frac{1}{2}+\alpha\bigr)
\Gamma\bigl(\frac{\beta}{2}\bigr)}\,
f_-\bigl(\alpha\bigr)g_-\bigl(\frac{\beta}{2}\bigr)\,,
\\[4pt]\label{c_+}
&&c_+(\alpha,\beta)=\frac{2^{\alpha-\beta}}
{\Gamma\bigl(\frac{1}{2}+\alpha-\frac{\beta}{2}\bigr)
\Gamma\bigl(-\frac{\beta}{2}\bigr)}\,
f_+\bigl(\alpha-\frac{\beta}{2}\bigr)g_+\bigl(-\frac{\beta}{2}\bigr)\,,
\\[4pt]\label{c_0}
&&c_0(\alpha)=\frac{2^{\alpha}}
{\Gamma\bigl(\frac{1}{2}+\alpha\bigr)}\,
f_0\bigl(\alpha\bigr)\,,
\end{eqnarray}
where the functions $f_s$, $g_s$, $s\in\{2;+;-;0\}$, are some 1-periodic
functions dependent on the initial data and $\arg x$. Now, we are ready to
describe the asymptotic behavior of the classical solutions of P4.

For example, let us consider the classical solutions corresponding to
the special points of the monodromy manifold (\ref{special_points}) where
$\alpha=\frac{1}{2}+n$, $n\in{\Bbb Z}_+$. Each solution of this series is
the B\"acklund transformation of the basic solution satisfying the Riccati
equation (\ref{ric_mm}) $y'+y^2+2xy+\beta=0$ of the form
$\Bigl(R_0^-\bigl(R_{\infty}^+\bigr)\Bigr)^n[y],$
since the superposition of the transformations $R_{\infty}^+$
(\ref{R+8_action}) and $R_0^-$ (\ref{R-0_action}) increases the value of the
parameter $\alpha$ in $1$ and preserves the parameter $\beta$. Starting from
(\ref{cy1}), we calculate the constant $f_-(\alpha)$ in
(\ref{c_-}) and obtain the asymptotics of the classical Painlev\'e function
for $\arg x\in\bigl(-\frac{\pi}{4};\frac{\pi}{4}\bigr)$.
Using the symmetry of the P4 equation (\ref{p4}) about rotation $x\mapsto ix$,
$y\mapsto iy$, $\alpha\mapsto-\alpha$, $\beta\mapsto-\beta$, we transfer the
asymptotics (\ref{cy2}) to the sector
$\bigl(-\frac{\pi}{4};\frac{\pi}{4}\bigr)$, obtain the product $f_2(p)g_2(q)$
in (\ref{c_2}) and then rotate back. Similar calculations yield result for
other sectors. Thus, the following assertion takes place:
\begin{theorem}\label{Tha=1/2+n}
For the parameter values
\begin{equation}\label{a=1/2+n}
\alpha=\frac{1}{2}+n,\quad
n\in{\Bbb Z}_+,\quad
\frac{\beta}{2}\notin{\Bbb Z}\,,
\end{equation}
there exist the classical solutions described asymptotically by the equations:
\begin{eqnarray}\label{sect0}
\hbox{i)}\quad
&&\arg x\in\bigl(-\frac{\pi}{4};\frac{\pi}{4}\bigr)\colon\nonumber
\\[4pt]
&&y=-\frac{\beta}{2x}\Bigl(1+{\cal O}\bigl(x^{-2}\bigr)\Bigr)+\mu_1
x^{2n+\beta}e^{-x^2}\Bigl(1+{\cal O}\bigl(x^{-2}\bigr)\Bigr),
\\[4pt]
&&\mu_1=-2\bigl(c+i\sqrt{\pi}\,\Theta(\arg x)\Bigr)
\frac{2^{n+\frac{\beta}{2}}}{n!\,\Gamma\bigl(\frac{\beta}{2}\bigr)},\quad
c\in{\Bbb C}\,;\nonumber
\\[4pt]
\hbox{ or }
&&y=-2x\Bigl(1+{\cal O}\bigl(x^{-2}\bigr)\Bigr),\quad
\hbox{ formally corresponding to }\quad
c=\infty;\nonumber
\\[8pt]\label{sect1}
\hbox{ii)}\quad
&&\arg x\in\bigl(\frac{\pi}{4};\frac{3\pi}{4}\bigr)\colon\nonumber
\\[4pt]
\hbox{if }\quad
&&c+i\sqrt{\pi}\neq0,\nonumber
\\[4pt]
\hbox{then }\quad
&&y=-2x\Bigl(1+{\cal O}\bigl(x^{-2}\bigr)\Bigr)+\mu_2
x^{4n+2-\beta}e^{x^2}\Bigl(1+{\cal O}\bigl(x^{-2}\bigr)\Bigr),
\\[4pt]
&&\mu_2=2\sqrt{\pi}\Bigl(\frac{i\sqrt{\pi}}{(c+i\sqrt{\pi})(e^{i\pi\beta}-1)}+
\Theta(\arg x-\frac{\pi}{2})\Bigr)
e^{i\frac{\pi\beta}{2}}\frac{2^{2n+1-\frac{\beta}{2}}}
{n!\,\Gamma\bigl(n+1-\frac{\beta}{2}\bigr)},\nonumber
\\[4pt]
\hbox{ and if }
&&c+i\sqrt{\pi}=0,\quad\hbox{ then }\quad
y=-\frac{\beta}{2x}\Bigl(1+{\cal O}\bigl(x^{-2}\bigr)\Bigr);\nonumber
\\[8pt]\label{sect2}
\hbox{iii)}\quad
&&\arg x\in\bigl(\frac{3\pi}{4};\frac{5\pi}{4}\bigr)\colon\nonumber
\\[4pt]
\hbox{if }\quad
&&c(1-e^{-i\pi\beta})+i\sqrt{\pi}\neq0,\nonumber
\\[4pt]
\hbox{then }\quad
&&y=-\frac{\beta}{2x}\Bigl(1+{\cal O}\bigl(x^{-2}\bigr)\Bigr)+\mu_3
x^{2n+\beta}e^{-x^2}\Bigl(1+{\cal O}\bigl(x^{-2}\bigr)\Bigr),
\\[4pt]
&&\mu_3=-2i\sqrt{\pi}
\Bigl(\frac{c+i\sqrt{\pi}}{c(1-e^{-i\pi\beta})+i\sqrt{\pi}}-
\Theta(\arg x-\pi)\Bigr)
e^{-i\pi\beta}
\frac{2^{n+\frac{\beta}{2}}}{n!\,\Gamma\bigl(\frac{\beta}{2}\bigr)},\nonumber
\\[4pt]
\hbox{ and if }
&&c(1-e^{-i\pi\beta})+i\sqrt{\pi}=0,\quad\hbox{ then }\quad
y=-2x\Bigl(1+{\cal O}\bigl(x^{-2}\bigr)\Bigr);\nonumber
\\[4pt]\label{sect3}
\hbox{iv)}\quad
&&\arg x\in\bigl(\frac{5\pi}{4};\frac{7\pi}{4}\bigr)\colon\nonumber
\\[4pt]
&&\hbox{if }\quad
c\neq0,\nonumber
\\[4pt]
\hbox{then }\quad
&&y=-2x\Bigl(1+{\cal O}\bigl(x^{-2}\bigr)\Bigr)+\mu_4
x^{4n+2-\beta}e^{x^2}\Bigl(1+{\cal O}\bigl(x^{-2}\bigr)\Bigr),
\\[4pt]
&&\mu_4=2\sqrt{\pi}\Bigl(\frac{i\sqrt{\pi}}{c(1-e^{-i\pi\beta})}+
\Theta(\frac{3\pi}{2}-\arg x)\Bigr)
e^{i\frac{3\pi\beta}{2}}
\frac{2^{2n+1-\frac{\beta}{2}}}{n!\,\Gamma\bigl(n+1-\frac{\beta}{2}\bigr)},
\nonumber
\\[4pt]
\hbox{ and if }\quad
&&c=0,\quad\hbox{ then }\quad
y=-\frac{\beta}{2x}\Bigl(1+{\cal O}\bigl(x^{-2}\bigr)\Bigr).\nonumber
\end{eqnarray}
\end{theorem}
To get the asymptotics of the classical Painlev\'e functions for
$\alpha=-\frac{1}{2}-n$, $n\in{\Bbb Z}_+$, it is enough to substitute $ix$,
$iy$, $-\alpha$, $-\beta$ instead of $x$, $y$, $\alpha$ and $\beta$,
respectively, in all the expressions above.

In the very similar way, starting from (\ref{ric_mp}) and taking into
account that the combination of the B\"acklund transformations
$R_{\infty}^{\pm}\bigl(R_{\infty}^{\pm}\bigr)$ changes the parameter value
$\beta$ in $2$ and preserves $\alpha$, we get the following assertions:
\begin{theorem}\label{Tha-b/2=n+1/2}
For the parameter values
\begin{equation}\label{a-b/2=n+1/2}
\alpha-\frac{\beta}{2}=\frac{1}{2}+n\,,\quad
n\in{\Bbb Z}_+,\quad
\alpha+\frac{1}{2}\notin{\Bbb Z}\,,
\end{equation}
there exist the classical solutions described asymptotically by the equations:
\begin{eqnarray}\label{a-b/2_sect0}
\hbox{i)}\quad
&&\arg x\in\bigl(-\frac{\pi}{4};\frac{\pi}{4}\bigr)\colon\nonumber
\\[4pt]
&&y=\frac{\beta}{2x}\Bigl(1+{\cal O}\bigl(x^{-2}\bigr)\Bigr)+\mu_1
x^{2n-\beta}e^{-x^2}\Bigl(1+{\cal O}\bigl(x^{-2}\bigr)\Bigr),
\\[4pt]
&&\mu_1=-2\bigl(c+i\sqrt{\pi}\,\Theta(\arg x)\Bigr)
\frac{2^{n-\frac{\beta}{2}}}
{n!\,\Gamma\bigl(-\frac{\beta}{2}\bigr)},\quad
c\in{\Bbb C}\,;\nonumber
\\[4pt]
\hbox{ or }
&&y=-2x\Bigl(1+{\cal O}\bigl(x^{-2}\bigr)\Bigr),\quad
\hbox{ formally corresponding to }\quad
c=\infty;\nonumber
\\[8pt]\label{a-b/2_sect1}
\hbox{ii)}\quad
&&\arg x\in\bigl(\frac{\pi}{4};\frac{3\pi}{4}\bigr)\colon\nonumber
\\[4pt]
\hbox{if }\quad
&&c+i\sqrt{\pi}\neq0,\nonumber
\\[4pt]
\hbox{then }\quad
&&y=-2x\Bigl(1+{\cal O}\bigl(x^{-2}\bigr)\Bigr)+\mu_2
x^{4\alpha-\beta}e^{x^2}\Bigl(1+{\cal O}\bigl(x^{-2}\bigr)\Bigr),
\\[4pt]
&&\mu_2=2\sqrt{\pi}\Bigl(
\frac{i\sqrt{\pi}}{(c+i\sqrt{\pi})(e^{-i\pi\beta}-1)}+
\Theta(\arg x-\frac{\pi}{2})\Bigr)
e^{-i\frac{\pi\beta}{2}}
\frac{2^{2\alpha-\frac{\beta}{2}}}
{n!\,\Gamma\bigl(\alpha+\frac{1}{2}\bigr)},\nonumber
\\[4pt]
\hbox{ and if }
&&c+i\sqrt{\pi}=0,\quad\hbox{ then }\quad
y=\frac{\beta}{2x}\Bigl(1+{\cal O}\bigl(x^{-2}\bigr)\Bigr);\nonumber
\\[8pt]\label{a-b/2_sect2}
\hbox{iii)}\quad
&&\arg x\in\bigl(\frac{3\pi}{4};\frac{5\pi}{4}\bigr)\colon\nonumber
\\[4pt]
\hbox{if }\quad
&&c(1-e^{i\pi\beta})+i\sqrt{\pi}\neq0,\nonumber
\\[4pt]
\hbox{then }\quad
&&y=\frac{\beta}{2x}\Bigl(1+{\cal O}\bigl(x^{-2}\bigr)\Bigr)+\mu_3
x^{2n-\beta}e^{-x^2}\Bigl(1+{\cal O}\bigl(x^{-2}\bigr)\Bigr),
\\[4pt]
&&\mu_3=-2i\sqrt{\pi}
\Bigl(\frac{c+i\sqrt{\pi}}{c(1-e^{i\pi\beta})+i\sqrt{\pi}}-
\Theta(\arg x-\pi)\Bigr)
e^{i\pi\beta}
\frac{2^{n-\frac{\beta}{2}}}
{n!\,\Gamma\bigl(-\frac{\beta}{2}\bigr)},\nonumber
\\[4pt]
\hbox{ and if }
&&c(1-e^{i\pi\beta})+i\sqrt{\pi}=0,
\quad\hbox{ then }\quad
y=-2x\Bigl(1+{\cal O}\bigl(x^{-2}\bigr)\Bigr);\nonumber
\\[4pt]\label{a-b/2_sect3}
\hbox{iv)}\quad
&&\arg x\in\bigl(\frac{5\pi}{4};\frac{7\pi}{4}\bigr)\colon\nonumber
\\[4pt]
\hbox{if }\quad
&&c\neq0,\nonumber
\\[4pt]
\hbox{then }\quad
&&y=-2x\Bigl(1+{\cal O}\bigl(x^{-2}\bigr)\Bigr)+\mu_4
x^{4\alpha-\beta}e^{x^2}\Bigl(1+{\cal O}\bigl(x^{-2}\bigr)\Bigr),
\\[4pt]
&&\mu_4=2\sqrt{\pi}\Bigl(\frac{i\sqrt{\pi}}{c(1-e^{i\pi\beta})}+
1-\Theta(\arg x-\frac{3\pi}{2})\Bigr)
e^{-i\frac{3\pi\beta}{2}}
\frac{2^{2\alpha-\frac{\beta}{2}}}
{n!\,\Gamma\bigl(\alpha+\frac{1}{2}\bigr)},
\nonumber
\\[4pt]
\hbox{ and if }\quad
&&c=0,\quad\hbox{ then }\quad
y=\frac{\beta}{2x}\Bigl(1+{\cal O}\bigl(x^{-2}\bigr)\Bigr).\nonumber
\end{eqnarray}
\end{theorem}
As before, to get the asymptotics of the classical Painlev\'e
functions for $\alpha-\beta/2=-n-1/2$, $n\in{\Bbb Z}_+$, it is enough to
substitute $ix$, $iy$, $-\alpha$, $-\beta$ instead of $x$, $y$, $\alpha$ and
$\beta$, respectively, in all the expressions above.

Next theorem follows by applying the B\"acklund transformation of the type
$R_{\infty}^{\pm}$ to the previous asymptotics:
\begin{theorem}\label{Thb=-2n}
For the parameter values
\begin{equation}\label{b=-2n}
\beta=-2n,\quad
n\in{\Bbb N}\,,\quad
\alpha+\frac{1}{2}\notin{\Bbb Z}\,,
\end{equation}
there exist the classical solutions described asymptotically by the equations:
\begin{eqnarray}\label{b-2n_sect0}
\hbox{i)}\quad
&&\arg x\in\bigl(-\frac{\pi}{4};\frac{\pi}{4}\bigr)\colon\nonumber
\\[4pt]
&&y=\frac{\beta}{2x}\Bigl(1+{\cal O}\bigl(x^{-2}\bigr)\Bigr)+\mu_1
x^{-1+4n+2\alpha}e^{-x^2}\Bigl(1+{\cal O}\bigl(x^{-2}\bigr)\Bigr),
\\[4pt]
&&\mu_1=-2\bigl(c+i\sqrt{\pi}\,\Theta(\arg x)\Bigr)
\frac{2^{-\frac{1}{2}+2n+\alpha}}
{(n-1)!\,\Gamma\bigl(\alpha+n+\frac{1}{2}\bigr)},\quad
c\in{\Bbb C}\,;\nonumber
\\[4pt]
\hbox{ or }
&&y=-\frac{\beta}{2x}\Bigl(1+{\cal O}\bigl(x^{-2}\bigr)\Bigr),\quad
\hbox{ formally corresponding to }\quad
c=\infty;\nonumber
\\[8pt]\label{b-2n_sect1}
\hbox{ii)}\quad
&&\arg x\in\bigl(\frac{\pi}{4};\frac{3\pi}{4}\bigr)\colon\nonumber
\\[4pt]
\hbox{if }\quad
&&c+i\sqrt{\pi}\neq0,\nonumber
\\[4pt]
\hbox{then }\quad
&&y=-\frac{\beta}{2x}\Bigl(1+{\cal O}\bigl(x^{-2}\bigr)\Bigr)+\mu_2
x^{2n-1-2\alpha}e^{x^2}\Bigl(1+{\cal O}\bigl(x^{-2}\bigr)\Bigr),
\\[4pt]
&&\mu_2=-2i\sqrt{\pi}\Bigl(
\frac{i\sqrt{\pi}}{(c+i\sqrt{\pi})(e^{2i\pi\alpha}+1)}-
\Theta(\arg x-\frac{\pi}{2})\Bigr)
\frac{(-1)^n\,2^{n-\frac{1}{2}-\alpha}e^{i\pi\alpha}}
{(n-1)!\,\Gamma\bigl(-\alpha+\frac{1}{2}\bigr)},\nonumber
\\[4pt]
\hbox{ and if }
&&c+i\sqrt{\pi}=0,\quad\hbox{ then }\quad
y=\frac{\beta}{2x}\Bigl(1+{\cal O}\bigl(x^{-2}\bigr)\Bigr);\nonumber
\\[8pt]\label{b-2n_sect2}
\hbox{iii)}\quad
&&\arg x\in\bigl(\frac{3\pi}{4};\frac{5\pi}{4}\bigr)\colon\nonumber
\\[4pt]
\hbox{if }\quad
&&c(e^{-2i\pi\alpha}+1)+i\sqrt{\pi}\neq0,\nonumber
\\[4pt]
\hbox{then }\quad
&&y=\frac{\beta}{2x}\Bigl(1+{\cal O}\bigl(x^{-2}\bigr)\Bigr)+\mu_3
x^{4n-1+2\alpha}e^{-x^2}\Bigl(1+{\cal O}\bigl(x^{-2}\bigr)\Bigr),
\\[4pt]
&&\hskip-.5cm
\mu_3=2i\sqrt{\pi}
\Bigl(\frac{c+i\sqrt{\pi}}{c(e^{-2i\pi\alpha}+1)+i\sqrt{\pi}}-
\Theta(\arg x-\pi)\Bigr)
\frac{2^{2n-\frac{1}{2}+\alpha}e^{-2i\pi\alpha}}
{(n-1)!\,\Gamma\bigl(\alpha+n+\frac{1}{2}\bigr)},\nonumber
\\[4pt]
\hbox{ and if }
&&c(e^{-2i\pi\alpha}+1)+i\sqrt{\pi}=0,\quad\hbox{ then }\quad
y=-\frac{\beta}{2x}\Bigl(1+{\cal O}\bigl(x^{-2}\bigr)\Bigr);\nonumber
\\[4pt]\label{b-2n_sect3}
\hbox{iv)}\quad
&&\arg x\in\bigl(\frac{5\pi}{4};\frac{7\pi}{4}\bigr)\colon\nonumber
\\[4pt]
\hbox{if }\quad
&&c\neq0,\nonumber
\\[4pt]
\hbox{then }\quad
&&y=-\frac{\beta}{2x}\Bigl(1+{\cal O}\bigl(x^{-2}\bigr)\Bigr)+\mu_4
x^{2n-1-2\alpha}e^{x^2}\Bigl(1+{\cal O}\bigl(x^{-2}\bigr)\Bigr),
\\[4pt]
&&\mu_4=-2i\sqrt{\pi}\Bigl(\frac{i\sqrt{\pi}}{c(e^{-2i\pi\alpha}+1)}+
1-\Theta(\arg x-\frac{3\pi}{2})\Bigr)
\frac{(-1)^n\,2^{n-\frac{1}{2}-\alpha}e^{3\pi i\alpha}}
{(n-1)!\,\Gamma\bigl(-\alpha+\frac{1}{2}\bigr)},
\nonumber
\\[4pt]
\hbox{ and if }\quad
&&c=0,\quad\hbox{ then }\quad
y=\frac{\beta}{2x}\Bigl(1+{\cal O}\bigl(x^{-2}\bigr)\Bigr).\nonumber
\end{eqnarray}
\end{theorem}
As before, to get the asymptotics of the classical Painlev\'e functions for
$\beta=2n$, $n\in{\Bbb N}\,$, it is enough to substitute $ix$,
$iy$, $-\alpha$, $-\beta$ instead of $x$, $y$, $\alpha$ and $\beta$,
respectively, in all the expressions above.

The last we are going to do in this section is the description of limiting
cases excluded from the theorems above. As easy to see, for any
half-integer $\alpha$ and even $\beta$, there are two different 1-parameter
families of the classical solutions. Namely,
\begin{theorem}\label{b-2m}
If
\begin{equation}\label{a=n+1/2,b=-2m}
\alpha=\frac{1}{2}+n\,,\quad
\beta=-2m,\quad
n,m\in{\Bbb Z}_+,
\end{equation}
then there exist two families of classical solutions:

1. The first family solutions are described by the asymptotics
\begin{eqnarray}\label{a=n+1/2,b=-2m_sect0}
\hbox{i)}\quad
&&\arg x\in\bigl(-\frac{\pi}{4};\frac{\pi}{4}\bigr)\colon\nonumber
\\[4pt]
&&y=-\frac{\beta}{2x}\Bigl(1+{\cal O}\bigl(x^{-2}\bigr)\Bigr)+\mu_1
x^{2n-2m}e^{-x^2}\Bigl(1+{\cal O}\bigl(x^{-2}\bigr)\Bigr),
\\[4pt]
&&\mu_1=-c(-1)^m\,2^{n-m+1}\,\frac{m!}{n!},\quad
c\in{\Bbb C}\,;\nonumber
\\[4pt]
\hbox{ or }
&&y=-2x\Bigl(1+{\cal O}\bigl(x^{-2}\bigr)\Bigr),\quad
\hbox{ formally corresponding to }\quad
c=\infty;\nonumber
\\[8pt]\label{a=n+1/2,b=-2m_sect1}
\hbox{ii)}\quad
&&\arg x\in\bigl(\frac{\pi}{4};\frac{3\pi}{4}\bigr)\colon\nonumber
\\[4pt]
\hbox{if }\quad
&&c\neq0,\nonumber
\\[4pt]
\hbox{then }\quad
&&y=-2x\Bigl(1+{\cal O}\bigl(x^{-2}\bigr)\Bigr)+\mu_2
x^{4n+2m+2}e^{x^2}\Bigl(1+{\cal O}\bigl(x^{-2}\bigr)\Bigr),
\\[4pt]
&&\mu_2=\Bigl(
\frac{1}{c}+2\sqrt{\pi}\,\Theta(\arg x-\frac{\pi}{2})\Bigr)
(-1)^m\,\frac{2^{2n+m+1}}{n!\,(n+m)!},\nonumber
\\[4pt]
\hbox{ and if }
&&c=0,\quad\hbox{ then }\quad
y=-\frac{\beta}{2x}\Bigl(1+{\cal O}\bigl(x^{-2}\bigr)\Bigr);\nonumber
\\[8pt]\label{a=n+1/2,b=-2m_sect2}
\hbox{iii)}\quad
&&\arg x\in\bigl(\frac{3\pi}{4};\frac{5\pi}{4}\bigr)\colon\nonumber
\\[4pt]
\hbox{if }\quad
&&1+2\sqrt{\pi}\,c\neq0,\nonumber
\\[4pt]
\hbox{then }\quad
&&y=-\frac{\beta}{2x}\Bigl(1+{\cal O}\bigl(x^{-2}\bigr)\Bigr)+\mu_3
x^{2n-2m}e^{-x^2}\Bigl(1+{\cal O}\bigl(x^{-2}\bigr)\Bigr),
\\[4pt]
&&\mu_3=-\frac{c}{1+2\sqrt{\pi}\,c}\,(-1)^m\,2^{n-m+1}\,
\frac{m!}{n!},\nonumber
\\[4pt]
\hbox{ and if }
&&1+2\sqrt{\pi}\,c=0,\quad\hbox{ then }\quad
y=-2x\Bigl(1+{\cal O}\bigl(x^{-2}\bigr)\Bigr);\nonumber
\\[4pt]\label{a=n+1/2,b=-2m_sect3}
\hbox{iv)}\quad
&&\arg x\in\bigl(\frac{5\pi}{4};\frac{7\pi}{4}\bigr)\colon\nonumber
\\[4pt]
\hbox{if }\quad
&&c\neq0,\nonumber
\\[4pt]
\hbox{then }\quad
&&y=-2x\Bigl(1+{\cal O}\bigl(x^{-2}\bigr)\Bigr)+\mu_4
x^{4n+2m+2}e^{x^2}\Bigl(1+{\cal O}\bigl(x^{-2}\bigr)\Bigr),
\\[4pt]
&&\mu_4=\Bigl(\frac{1}{c}+2\sqrt{\pi}\,\Theta(\frac{3\pi}{2}-\arg x)\Bigr)
(-1)^m\,\frac{2^{2n+m+1}}{n!\,(n+m)!},
\nonumber
\\[4pt]
\hbox{ and if }\quad
&&c=0,\quad\hbox{ then }\quad
y=-\frac{\beta}{2x}\Bigl(1+{\cal O}\bigl(x^{-2}\bigr)\Bigr).\nonumber
\end{eqnarray}
2. Solutions of the second family behave asymptotically as follows:
\begin{eqnarray}\label{a=n+1/2,b=-2m_sect0a}
\hbox{i)}\quad
&&\arg x\in\bigl(-\frac{\pi}{4};\frac{\pi}{4}\bigr)\colon\nonumber
\\[4pt]
&&y=\frac{\beta}{2x}\Bigl(1+{\cal O}\bigl(x^{-2}\bigr)\Bigr)+\mu_1
x^{2n+4m}e^{-x^2}\Bigl(1+{\cal O}\bigl(x^{-2}\bigr)\Bigr),
\\[4pt]
&&\mu_1=-\Bigl(c+i\sqrt{\pi}\,\Theta(\arg x)\Bigr)
\frac{2^{n+2m+1}}{(n+m)!\,(m-1)!},\quad
c\in{\Bbb C}\,;\nonumber
\\[4pt]
\hbox{ or }
&&y=-\frac{\beta}{2x}\Bigl(1+{\cal O}\bigl(x^{-2}\bigr)\Bigr),\quad
\hbox{ formally corresponding to }\quad
c=\infty;\nonumber
\\[8pt]\label{a=n+1/2,b=-2m_sect1a}
\hbox{ii)}\quad
&&\arg x\in\bigl(\frac{\pi}{4};\frac{3\pi}{4}\bigr)\colon\nonumber
\\[4pt]
\hbox{if }\quad
&&c+i\sqrt{\pi}\neq0,\nonumber
\\[4pt]
\hbox{then }\quad
&&y=-\frac{\beta}{2x}\Bigl(1+{\cal O}\bigl(x^{-2}\bigr)\Bigr)+\mu_2
x^{2m-2n-2}e^{x^2}\Bigl(1+{\cal O}\bigl(x^{-2}\bigr)\Bigr),
\\[4pt]
&&\mu_2=\frac{1}{c+i\sqrt{\pi}}\,(-1)^m\,2^{m-n-1}\frac{n!}{(m-1)!},\nonumber
\\[4pt]
\hbox{ and if }
&&c+i\sqrt{\pi}=0,\quad\hbox{ then }\quad
y=\frac{\beta}{2x}\Bigl(1+{\cal O}\bigl(x^{-2}\bigr)\Bigr);\nonumber
\\[8pt]\label{a=n+1/2,b=-2m_sect2a}
\hbox{iii)}\quad
&&\arg x\in\bigl(\frac{3\pi}{4};\frac{5\pi}{4}\bigr)\colon\nonumber
\\[4pt]
\hbox{if }\quad
&&c\neq\infty,\nonumber
\\[4pt]
\hbox{then }\quad
&&y=\frac{\beta}{2x}\Bigl(1+{\cal O}\bigl(x^{-2}\bigr)\Bigr)+\mu_3
x^{2n+4m}e^{-x^2}\Bigl(1+{\cal O}\bigl(x^{-2}\bigr)\Bigr),
\\[4pt]
&&\mu_3=-\Bigl(c+i\sqrt{\pi}\,\Theta(\pi-\arg x)\Bigr)
\frac{2^{n+2m+1}}{(n+m)!\,(m-1)!},\nonumber
\\[4pt]
\hbox{ and if }
&&c=\infty,\quad\hbox{ then }\quad
y=-\frac{\beta}{2x}\Bigl(1+{\cal O}\bigl(x^{-2}\bigr)\Bigr);\nonumber
\\[4pt]\label{a=n+1/2,b=-2m_sect3a}
\hbox{iv)}\quad
&&\arg x\in\bigl(\frac{5\pi}{4};\frac{7\pi}{4}\bigr)\colon\nonumber
\\[4pt]
\hbox{if }\quad
&&c\neq0,\nonumber
\\[4pt]
\hbox{then }\quad
&&y=-\frac{\beta}{2x}\Bigl(1+{\cal O}\bigl(x^{-2}\bigr)\Bigr)+\mu_4
x^{2m-2n-2}e^{x^2}\Bigl(1+{\cal O}\bigl(x^{-2}\bigr)\Bigr),
\\[4pt]
&&\mu_4=\frac{1}{c}\,(-1)^m\,2^{m-n-1}\frac{n!}{(m-1)!},\nonumber
\\[4pt]
\hbox{ and if }\quad
&&c=0,\quad\hbox{ then }\quad
y=\frac{\beta}{2x}\Bigl(1+{\cal O}\bigl(x^{-2}\bigr)\Bigr).\nonumber
\end{eqnarray}
In this case, the value $m=0$ corresponds to the trivial solution
$y\equiv0$.
\end{theorem}

\begin{theorem}\label{b2m<n+1}
If
\begin{equation}\label{a=n+1/2,b=2m<n+1}
\alpha=\frac{1}{2}+n\,,\quad
\beta=2m,\quad
n,m\in{\Bbb Z}_+,\quad
m\leq n\,,
\end{equation}
then there exist two families of the classical solutions:

1. The solutions of the first family are described by the asymptotic
relations:
\begin{eqnarray}\label{a=n+1/2,b=2m<n+1_sect0}
\hbox{i)}\quad
&&\arg x\in\bigl(-\frac{\pi}{4};\frac{\pi}{4}\bigr)\colon\nonumber
\\[4pt]
&&y=\frac{\beta}{2x}\Bigl(1+{\cal O}\bigl(x^{-2}\bigr)\Bigr)+\mu_1
x^{2n-4m}e^{-x^2}\Bigl(1+{\cal O}\bigl(x^{-2}\bigr)\Bigr),
\\[4pt]
&&\mu_1=-c(-1)^m\,2^{n-2m+1}\frac{m!}{(n-m)!},\quad
c\in{\Bbb C}\,;\nonumber
\\[4pt]
\hbox{ or }
&&y=-2x\Bigl(1+{\cal O}\bigl(x^{-2}\bigr)\Bigr),\quad
\hbox{ formally corresponding to }\quad
c=\infty;\nonumber
\\[8pt]\label{a=n+1/2,b=2m<n+1_sect1}
\hbox{ii)}\quad
&&\arg x\in\bigl(\frac{\pi}{4};\frac{3\pi}{4}\bigr)\colon\nonumber
\\[4pt]
\hbox{if }\quad
&&c\neq0,\nonumber
\\[4pt]
\hbox{then }\quad
&&y=-2x\Bigl(1+{\cal O}\bigl(x^{-2}\bigr)\Bigr)+\mu_2
x^{4n-2m+2}e^{x^2}\Bigl(1+{\cal O}\bigl(x^{-2}\bigr)\Bigr),
\\[4pt]
&&\mu_2=\Bigl(
\frac{1}{c}+2\sqrt{\pi}\,\Theta(\arg x-\frac{\pi}{2})\Bigr)
(-1)^m\,\frac{2^{2n-m+1}}{n!\,(n-m)!},\nonumber
\\[4pt]
\hbox{ and if }
&&c=0,\quad\hbox{ then }\quad
y=\frac{\beta}{2x}\Bigl(1+{\cal O}\bigl(x^{-2}\bigr)\Bigr);\nonumber
\\[8pt]\label{a=n+1/2,b=2m<n+1_sect2}
\hbox{iii)}\quad
&&\arg x\in\bigl(\frac{3\pi}{4};\frac{5\pi}{4}\bigr)\colon\nonumber
\\[4pt]
\hbox{if }\quad
&&1+2\sqrt{\pi}\,c\neq0,\nonumber
\\[4pt]
\hbox{then }\quad
&&y=\frac{\beta}{2x}\Bigl(1+{\cal O}\bigl(x^{-2}\bigr)\Bigr)+\mu_3
x^{2n-4m}e^{-x^2}\Bigl(1+{\cal O}\bigl(x^{-2}\bigr)\Bigr),
\\[4pt]
&&\mu_3=-\frac{c}{1+2\sqrt{\pi}\,c}(-1)^m\,2^{n-2m+1}\,
\frac{m!}{(n-m)!},\nonumber
\\[4pt]
\hbox{ and if }
&&1+2\sqrt{\pi}\,c=0,\quad\hbox{ then }\quad
y=-2x\Bigl(1+{\cal O}\bigl(x^{-2}\bigr)\Bigr);\nonumber
\\[4pt]\label{a=n+1/2,b=2m<n+1_sect3}
\hbox{iv)}\quad
&&\arg x\in\bigl(\frac{5\pi}{4};\frac{7\pi}{4}\bigr)\colon\nonumber
\\[4pt]
\hbox{if }\quad
&&c\neq0,\nonumber
\\[4pt]
\hbox{then }\quad
&&y=-2x\Bigl(1+{\cal O}\bigl(x^{-2}\bigr)\Bigr)+\mu_4
x^{4n-2m+2}e^{x^2}\Bigl(1+{\cal O}\bigl(x^{-2}\bigr)\Bigr),
\\[4pt]
&&\mu_4=\Bigl(\frac{1}{c}+2\sqrt{\pi}\,
\Theta(\frac{3\pi}{2}-\arg x)\Bigr)(-1)^m\,\frac{2^{2n-m+1}}{n!\,(n-m)!},
\nonumber
\\[4pt]
\hbox{ and if }\quad
&&c=0,\quad\hbox{ then }\quad
y=\frac{\beta}{2x}\Bigl(1+{\cal O}\bigl(x^{-2}\bigr)\Bigr).\nonumber
\end{eqnarray}
2. The solutions of the second family are as follows:
\begin{eqnarray}\label{a=n+1/2,b=2m<n+1_sect0a}
\hbox{i)}\quad
&&\arg x\in\bigl(-\frac{\pi}{4};\frac{\pi}{4}\bigr)\colon\nonumber
\\[4pt]
&&y=-\frac{\beta}{2x}\Bigl(1+{\cal O}\bigl(x^{-2}\bigr)\Bigr)+\mu_1
x^{2n+2m}e^{-x^2}\Bigl(1+{\cal O}\bigl(x^{-2}\bigr)\Bigr),
\\[4pt]
&&\mu_1=-\Bigl(c+i\sqrt{\pi}\,\Theta(\arg x)\Bigr)
\frac{2^{n+m+1}}{n!\,(m-1)!},\quad
c\in{\Bbb C}\,;\nonumber
\\[4pt]
\hbox{ or }
&&y=\frac{\beta}{2x}\Bigl(1+{\cal O}\bigl(x^{-2}\bigr)\Bigr),\quad
\hbox{ formally corresponding to }\quad
c=\infty;\nonumber
\\[8pt]\label{a=n+1/2,b=2m<n+1_sect1a}
\hbox{ii)}\quad
&&\arg x\in\bigl(\frac{\pi}{4};\frac{3\pi}{4}\bigr)\colon\nonumber
\\[4pt]
\hbox{if }\quad
&&c+i\sqrt{\pi}\neq0,\nonumber
\\[4pt]
\hbox{then }\quad
&&y=\frac{\beta}{2x}\Bigl(1+{\cal O}\bigl(x^{-2}\bigr)\Bigr)+\mu_2
x^{4m-2n-2}e^{x^2}\Bigl(1+{\cal O}\bigl(x^{-2}\bigr)\Bigr),
\\[4pt]
&&\mu_2=\frac{1}{c+i\sqrt{\pi}}\,
(-1)^m\,2^{2m-n-1}\frac{(n-m)!}{(m-1)!},\nonumber
\\[4pt]
\hbox{ and if }
&&c+i\sqrt{\pi}=0,\quad\hbox{ then }\quad
y=-\frac{\beta}{2x}\Bigl(1+{\cal O}\bigl(x^{-2}\bigr)\Bigr);\nonumber
\\[8pt]\label{a=n+1/2,b=2m<n+1_sect2a}
\hbox{iii)}\quad
&&\arg x\in\bigl(\frac{3\pi}{4};\frac{5\pi}{4}\bigr)\colon\nonumber
\\[4pt]
\hbox{if }\quad
&&c\neq\infty,\nonumber
\\[4pt]
\hbox{then }\quad
&&y=-\frac{\beta}{2x}\Bigl(1+{\cal O}\bigl(x^{-2}\bigr)\Bigr)+\mu_3
x^{2n+2m}e^{-x^2}\Bigl(1+{\cal O}\bigl(x^{-2}\bigr)\Bigr),
\\[4pt]
&&\mu_3=-\Bigl(c+i\sqrt{\pi}\,\Theta(\pi-\arg x)\Bigr)
\frac{2^{n+m+1}}{n!\,(m-1)!},\nonumber
\\[4pt]
\hbox{ and if }
&&c=\infty,\quad\hbox{ then }\quad
y=\frac{\beta}{2x}\Bigl(1+{\cal O}\bigl(x^{-2}\bigr)\Bigr);\nonumber
\\[4pt]\label{a=n+1/2,b=2m<n+1_sect3a}
\hbox{iv)}\quad
&&\arg x\in\bigl(\frac{5\pi}{4};\frac{7\pi}{4}\bigr)\colon\nonumber
\\[4pt]
\hbox{if }\quad
&&c\neq0,\nonumber
\\[4pt]
\hbox{then }\quad
&&y=\frac{\beta}{2x}\Bigl(1+{\cal O}\bigl(x^{-2}\bigr)\Bigr)+\mu_4
x^{4m-2n-2}e^{x^2}\Bigl(1+{\cal O}\bigl(x^{-2}\bigr)\Bigr),
\\[4pt]
&&\mu_4=\frac{1}{c}\,(-1)^m\,2^{2m-n-1}\frac{(n-m)!}{(m-1)!},
\nonumber
\\[4pt]
\hbox{ and if }\quad
&&c=0,\quad\hbox{ then }\quad
y=-\frac{\beta}{2x}\Bigl(1+{\cal O}\bigl(x^{-2}\bigr)\Bigr).\nonumber
\end{eqnarray}
In this case, the value $m=0$ corresponds to the trivial solution
$y\equiv0$.
\end{theorem}

\begin{theorem}\label{b2m>n}
If
\begin{equation}\label{a=n+1/2,b=2m>n}
\alpha=\frac{1}{2}+n\,,\quad
\beta=2m,\quad
n,m\in{\Bbb Z}_+,\quad
m\geq n+1\,,
\end{equation}
then there exist two families of the classical solutions:

1. The solutions of the first family are given by the asymptotic relations:
\begin{eqnarray}\label{a=n+1/2,b=2m>n_sect0}
\hbox{i)}\quad
&&\arg x\in\bigl(-\frac{\pi}{4};\frac{\pi}{4}\bigr)\colon\nonumber
\\[4pt]
&&y=-\frac{\beta}{2x}\Bigl(1+{\cal O}\bigl(x^{-2}\bigr)\Bigr)+\mu_1
x^{2n+2m}e^{-x^2}\Bigl(1+{\cal O}\bigl(x^{-2}\bigr)\Bigr),
\\[4pt]
&&\mu_1=-\Bigl(c+i\sqrt{\pi}\,\Theta(\arg x)\Bigr)
\frac{2^{n+m+1}}{n!\,(m-1)!},\quad
c\in{\Bbb C}\,;\nonumber
\\[4pt]
\hbox{ or }
&&y=-2x\Bigl(1+{\cal O}\bigl(x^{-2}\bigr)\Bigr),\quad
\hbox{ formally corresponding to }\quad
c=\infty;\nonumber
\\[8pt]\label{a=n+1/2,b=2m>n_sect1}
\hbox{ii)}\quad
&&\arg x\in\bigl(\frac{\pi}{4};\frac{3\pi}{4}\bigr)\colon\nonumber
\\[4pt]
\hbox{if }\quad
&&c+i\sqrt{\pi}\neq0,\nonumber
\\[4pt]
\hbox{then }\quad
&&y=-2x\Bigl(1+{\cal O}\bigl(x^{-2}\bigr)\Bigr)+\mu_2
x^{4n-2m+2}e^{x^2}\Bigl(1+{\cal O}\bigl(x^{-2}\bigr)\Bigr),
\\[4pt]
&&\mu_2=\frac{1}{c+i\sqrt{\pi}}\,(-1)^n\,
2^{2n-m+1}\frac{(m-n-1)!}{n!},\nonumber
\\[4pt]
\hbox{ and if }
&&c+i\sqrt{\pi}=0,\quad\hbox{ then }\quad
y=-\frac{\beta}{2x}\Bigl(1+{\cal O}\bigl(x^{-2}\bigr)\Bigr);\nonumber
\\[8pt]\label{a=n+1/2,b=2m>n_sect2}
\hbox{iii)}\quad
&&\arg x\in\bigl(\frac{3\pi}{4};\frac{5\pi}{4}\bigr)\colon\nonumber
\\[4pt]
\hbox{if }\quad
&&c\neq\infty,\nonumber
\\[4pt]
\hbox{then }\quad
&&y=-\frac{\beta}{2x}\Bigl(1+{\cal O}\bigl(x^{-2}\bigr)\Bigr)+\mu_3
x^{2n+2m}e^{-x^2}\Bigl(1+{\cal O}\bigl(x^{-2}\bigr)\Bigr),
\\[4pt]
&&\mu_3=-\Bigl(c+i\sqrt{\pi}\,\Theta(\pi-\arg x)\Bigr)
\frac{2^{n+m+1}}{n!\,(m-1)!},\nonumber
\\[4pt]
\hbox{ and if }
&&c=\infty,\quad\hbox{ then }\quad
y=-2x\Bigl(1+{\cal O}\bigl(x^{-2}\bigr)\Bigr);\nonumber
\\[4pt]\label{a=n+1/2,b=2m>n_sect3}
\hbox{iv)}\quad
&&\arg x\in\bigl(\frac{5\pi}{4};\frac{7\pi}{4}\bigr)\colon\nonumber
\\[4pt]
\hbox{if }\quad
&&c\neq0,\nonumber
\\[4pt]
\hbox{then }\quad
&&y=-2x\Bigl(1+{\cal O}\bigl(x^{-2}\bigr)\Bigr)+\mu_4
x^{4n-2m+2}e^{x^2}\Bigl(1+{\cal O}\bigl(x^{-2}\bigr)\Bigr),
\\[4pt]
&&\mu_4=\frac{1}{c}\,(-1)^n\,2^{2n-m+1}\frac{(m-n-1)!}{n!},
\nonumber
\\[4pt]
\hbox{ and if }\quad
&&c=0,\quad\hbox{ then }\quad
y=-\frac{\beta}{2x}\Bigl(1+{\cal O}\bigl(x^{-2}\bigr)\Bigr).\nonumber
\end{eqnarray}
2. The solutions of the second family are as follows:
\begin{eqnarray}\label{a=n+1/2,b=2m>n_sect0a}
\hbox{i)}\quad
&&\arg x\in\bigl(-\frac{\pi}{4};\frac{\pi}{4}\bigr)\colon\nonumber
\\[4pt]
&&y=-2x\Bigl(1+{\cal O}\bigl(x^{-2}\bigr)\Bigr)+\mu_1
x^{2m-4n-2}e^{-x^2}\Bigl(1+{\cal O}\bigl(x^{-2}\bigr)\Bigr),
\\[4pt]
&&\mu_1=c\,(-1)^n\,2^{m-2n}\frac{n!}{(m-n-1)!},\quad
c\in{\Bbb C}\,,\nonumber
\\[4pt]
\hbox{ or }
&&y=\frac{\beta}{2x}\Bigl(1+{\cal O}\bigl(x^{-2}\bigr)\Bigr),\quad
\hbox{ formally corresponding to }\quad
c=\infty;\nonumber
\\[8pt]\label{a=n+1/2,b=2m>n_sect1a}
\hbox{ii)}\quad
&&\arg x\in\bigl(\frac{\pi}{4};\frac{3\pi}{4}\bigr)\colon\nonumber
\\[4pt]
\hbox{if }\quad
&&c\neq0,\nonumber
\\[4pt]
\hbox{then }\quad
&&y=\frac{\beta}{2x}\Bigl(1+{\cal O}\bigl(x^{-2}\bigr)\Bigr)+\mu_2
x^{4m-2n-2}e^{x^2}\Bigl(1+{\cal O}\bigl(x^{-2}\bigr)\Bigr),
\\[4pt]
&&\mu_2=-\Bigl(\frac{1}{c}+2\sqrt{\pi}\,\Theta(\arg x-\frac{\pi}{2})\Bigr)
(-1)^n\,\frac{2^{2m-n-1}}{(m-1)!\,(m-n-1)!},\nonumber
\\[4pt]
\hbox{ and if }
&&c=0,\quad\hbox{ then }\quad
y=-2x\Bigl(1+{\cal O}\bigl(x^{-2}\bigr)\Bigr);\nonumber
\\[8pt]\label{a=n+1/2,b=2m>n_sect2a}
\hbox{iii)}\quad
&&\arg x\in\bigl(\frac{3\pi}{4};\frac{5\pi}{4}\bigr)\colon\nonumber
\\[4pt]
\hbox{if }\quad
&&1+2\sqrt{\pi}\,c\neq0,\nonumber
\\[4pt]
\hbox{then }\quad
&&y=-2x\Bigl(1+{\cal O}\bigl(x^{-2}\bigr)\Bigr)+\mu_3
x^{2m-4n-2}e^{-x^2}\Bigl(1+{\cal O}\bigl(x^{-2}\bigr)\Bigr),
\\[4pt]
&&\mu_3=\frac{c}{1+2\sqrt{\pi}\,c}\,(-1)^n\,
2^{m-2n}\frac{n!}{(m-n-1)!},\nonumber
\\[4pt]
\hbox{ and if }
&&1+2\sqrt{\pi}\,c=0,\quad\hbox{ then }\quad
y=\frac{\beta}{2x}\Bigl(1+{\cal O}\bigl(x^{-2}\bigr)\Bigr);\nonumber
\\[4pt]\label{a=n+1/2,b=2m>n_sect3a}
\hbox{iv)}\quad
&&\arg x\in\bigl(\frac{5\pi}{4};\frac{7\pi}{4}\bigr)\colon\nonumber
\\[4pt]
\hbox{if }\quad
&&c\neq0,\nonumber
\\[4pt]
\hbox{then }\quad
&&y=\frac{\beta}{2x}\Bigl(1+{\cal O}\bigl(x^{-2}\bigr)\Bigr)+\mu_4
x^{4m-2n-2}e^{x^2}\Bigl(1+{\cal O}\bigl(x^{-2}\bigr)\Bigr),
\\[4pt]
&&\mu_4=-\Bigl(\frac{1}{c}+2\sqrt{\pi}\,
\Theta\bigl(\frac{3\pi}{2}-\arg x\bigr)\Bigr)\,(-1)^n\,
\frac{2^{2m-n-1}}{(m-n-1)!\,(m-1)!},\nonumber
\\[4pt]
\hbox{ and if }\quad
&&c=0,\quad\hbox{ then }\quad
y=-2x\Bigl(1+{\cal O}\bigl(x^{-2}\bigr)\Bigr).\nonumber
\end{eqnarray}
\end{theorem}

As easy to see, Theorems~\ref{b-2m}, \ref{b2m<n+1} and~\ref{b2m>n} allow
some classical solutions to have uniform asymptotics of the type
$\pm\beta/2x$ or $-2x$ on the $x$ complex plane. In fact, these special
cases correspond to the rational solutions of the equation P4 described
above. Furthermore, using these theorems, we can specify the parameters
values which allow the rational solutions to exist:

\noindent
a) the parameters $\alpha=\pm\bigl(n+\frac{1}{2}\bigr)$, $\beta=\mp2m$,
$n,m\in{\Bbb Z}_+$, allow the rational solutions with the asymptotics
$y=-\frac{\beta}{2x}+{\cal O}\bigl(x^{-3}\bigr)=\pm\frac{m}{x}+
{\cal O}\bigl(x^{-3}\bigr)$ to exist;

\noindent
b) the parameters $\alpha=\pm\bigl(n+\frac{1}{2}\bigr)$, $\beta=\pm2m$,
$n,m\in{\Bbb Z}_+$, $m\leq n$, allow the rational solutions with the
asymptotics $y=\frac{\beta}{2x}+{\cal O}\bigl(x^{-3}\bigr)=\pm\frac{m}{x}+
{\cal O}\bigl(x^{-3}\bigr)$ to exist;

\noindent
c) the parameters $\alpha=\pm\bigl(n+\frac{1}{2}\bigr)$, $\beta=\pm2m$,
$n,m\in{\Bbb Z}_+$, $m\geq n+1$, allow the rational solutions with the
asymptotics $y=-2x+{\cal O}\bigl(x^{-1}\bigr)$ to exist.

\section{Transcendent degenerated Painlev\'e functions}\label{connection}

All the classical solutions are {\it truncated} in the interior of all of the
complex sectors $\arg x\in\bigl(\frac{\pi}{4}(m-1)+\frac{\pi}{2}n\,;\
\frac{\pi}{4}m+\frac{\pi}{2}n\bigr)$, $\forall n,m\colon$ $n\in{\Bbb Z}\,$,
$m\in\{0;1\}$. In fact, this is the characteristic property of the classical
Painlev\'e functions. Indeed, it is shown in the work \cite{Kap} that in the
case of general position distinguished by the inequality
$s_{2+n}\bigl(s_{1+n}+s_{3+n}+s_{1+n}s_{2+n}s_{3+n}\bigr)
\bigl(1+s_{1+n+m}s_{2+n+m}\bigr)\neq0\,,$
the Painlev\'e function is described asymptotically by means of some elliptic
function of the periods depending on $\arg x$ in the interior of the indicated
sector. Hence, the condition for the non-elliptic asymptotic behavior
in the interior of any such sector is reduced to the system
$$
s_{2+n}\bigl(s_{1+n}+s_{3+n}+s_{1+n}s_{2+n}s_{3+n}\bigr)
\bigl(1+s_{1+n+m}s_{2+n+m}\bigr)=0\quad
\forall n,m\colon\
n\in{\Bbb Z}\,,\
m\in\{0;1\}.$$
Direct checking shows that the last system of equations leads either to
the constraint (\ref{special_points}), or to (\ref{sk_rat_2/3}), or to
$s_1=s_3=0$, or to $s_2=s_4=0$, any of which corresponds to a classical
Painlev\'e function (see above).

In contrast to the case of the classical solutions, the Painlev\'e function
which has the elliptic asymptotics inside at least one of the sectors can not
be expressed via the classical special functions and should be called
transcendent. The function is parameterized by the Stokes multipliers $s_k$,
and the monodromy surface equation (\ref{monodromy_surface}) with the
symmetry (\ref{s_k+4}) together yields the connection formulae for it.

Besides the asymptotic solutions of general position for any $\arg x$, the
work \cite{Kap} contains description of some transcendent degenerated cases
corresponding to the following Stokes multiplier values (for simplicity, we
omit here the number shift $n$): 1)~$s_2=0$, $s_1+s_3\neq0$; 2)~$s_2\neq0$,
$s_1+s_3+s_1s_2s_3=0$, $\frac{1}{2}+\alpha\notin{\Bbb Z}\,$; and
3)~$1+s_2s_3=0$. Below, the rest of
the transcendent degenerated cases is described. As in ref.\ \cite{Kap}, the
results are formulated for any of the opened sectors
$\arg x\in\bigl(-\frac{\pi}{4}+\frac{\pi}{2}n;
\frac{\pi}{4}+\frac{\pi}{2}n\bigr)$, $n\in{\Bbb Z}\,$.

Let us begin with the case
\begin{equation}\label{restriction_sk}
s_2=s_1+s_3=0\,.
\end{equation}
The equation (\ref{restriction_sk}) with the equation of the monodromy
surface (\ref{monodromy_surface}) imply the restriction
$\cos\pi\bigl(\alpha-\frac{\beta}{2}\bigr)\sin\pi\frac{\beta}{2}=0$,
i.e.\ $\beta$ is even or $2\alpha-\beta$ is odd. Theorem~\ref{Theorem1} below
is obtained in the paper \cite{IK} for the case (\ref{restriction_sk}) with
the additional conditions $\beta=0$ and $\alpha+\frac{1}{2}\notin{\Bbb N}\,$ for
$x\to+\infty$. Below, using the symmetry (\ref{s_symmetry_rotation}),
(\ref{y_symmetry_rotation}) we give its elementary generalization.

\begin{theorem}\label{Theorem1} If $x\to\infty$,\quad
$\arg (x)\in\bigl(-\frac{\pi}{4}+\frac{\pi}{2}\,n;
\frac{\pi}{4}+\frac{\pi}{2}\,n\bigr)$ for some $n\in {\Bbb Z}\,$, and
\begin{equation}\label{trivial_sk1}
s_{2+n}=s_{1+n}+s_{3+n}=0,\quad
\beta=0,\quad (-1)^n\,\alpha+\frac{1}{2}\notin{\Bbb N}\,,
\end{equation}
then the corresponding solution $y(x)$ of the
fourth Painlev\'e equation P4 possesses the following asymptotic
behavior:
\begin{eqnarray}\label{IK_formula}
&&y=(-1)^n\,\frac{s_{1+n}s_{4+n}}{\pi^{3/2}}\,
e^{-i\pi(1+n)(-1)^n\alpha}
\Gamma\Bigl(\frac{1}{2}-(-1)^n\alpha\Bigr)
2^{(-1)^n\alpha-\frac{3}{2}}\times\nonumber
\\[4pt]
&&\times
x^{2(-1)^n\alpha-1}e^{-(-1)^nx^2}
\bigl( 1+{\cal O}(x^{-1})\bigr)\, .
\end{eqnarray}
\end{theorem}

Theorem~\ref{Theorem1} is proved in \cite{IK} by means of direct asymptotic
investigation of the Riemann-Hilbert problem for the function $\Psi(\lambda)$
and affirms the existence of the genuine Painlev\'e function described
asymptotically via the formal expression (\ref{formal_form}), (\ref{y_0}) for
$y_0$ with the main term (\ref{IK_formula}) depending on the Stokes
multipliers. Next Theorem~\ref{Theorem2} follows from Theorem~\ref{Theorem1}
after applying the B\"acklund transformations
(\ref{backlund_0^+})--(\ref{backlund_infty^-}) in the way described in the
previous section.

\begin{theorem}\label{Theorem2} Let $x\to\infty$,\quad
$\arg (x)\in\bigl(-\frac{\pi}{4}+\frac{\pi}{2}\,n;
\frac{\pi}{4}+\frac{\pi}{2}\,n\bigr)$ for some
$n\in {\Bbb Z}\,$, and
\begin{equation}\label{trivial_sk2}
s_{2+n}=s_{1+n}+s_{3+n}=0.
\end{equation}
\begin{description}
\item{(i)} Let
\begin{equation}\label{ab2i}
(-1)^n\beta=2k,\quad k\in{\Bbb Z}_+\,,\quad
(-1)^n\alpha+\frac{1}{2}-k\notin{\Bbb N}\,,
\end{equation}
then the corresponding solution $y(x)$ of the
fourth Painlev\'e equation P4 possesses the following asymptotic
behavior:
\begin{eqnarray}\label{asymptotics2i}
&&y=e^{i\frac{\pi}{2}n}y_+(e^{-i\frac{\pi}{2}n}x,(-1)^na,(-1)^n\beta,c_n)=
\nonumber
\\[4pt]
&&=\frac{\beta}{2x}+\frac{\beta(2\alpha-\frac{3\beta}{2})}{4x^3}+
{\cal O}\bigl(\frac{1}{x^5}\bigr)+
\\[4pt]
&&+(-1)^n\,\frac{s_{1+n}s_{4+n}}{\pi^{3/2}}\,
e^{-i\pi(n+1)(-1)^n\alpha}\,
\Gamma\bigl(\frac{1}{2}+k-(-1)^n\alpha\bigr)\times\nonumber
\\[4pt]
&&\times
2^{(-1)^n\alpha-\frac{3}{2}-2k}\,k!\,
x^{2(-1)^n\alpha-1-4k}e^{-(-1)^nx^2}
\bigl( 1+{\cal O}(x^{-2})\bigr)\, ;\nonumber
\end{eqnarray}
\item{(ii)} Let
\begin{equation}\label{ab2ii}
(-1)^n\beta=-2k,\quad k\in{\Bbb Z}_+\,,\quad
(-1)^n\alpha+\frac{1}{2}\notin{\Bbb N}\,,
\end{equation}
then the corresponding solution $y(x)$ of the
fourth Painlev\'e equation P4 possesses the following asymptotic
behavior:
\begin{eqnarray}\label{asymptotics2ii}
&&y=e^{i\frac{\pi}{2}n}y_-(e^{-i\frac{\pi}{2}n}x,(-1)^na,(-1)^n\beta,c_n)=
\nonumber
\\[4pt]
&&=-\frac{\beta}{2x}-\frac{\beta(2\alpha+\frac{\beta}{2})}{4x^3}+
{\cal O}\bigl(\frac{1}{x^5}\bigr)+
\\[4pt]
&&+(-1)^{n+k+nk}\,\frac{s_{1+n}s_{4+n}}{\pi^{3/2}}\,
e^{-i\pi(n+1)(-1)^n\alpha}\,
\Gamma\bigl(\frac{1}{2}-(-1)^n\alpha\bigr)\times\nonumber
\\[4pt]
&&\times 2^{(-1)^n\alpha-\frac{3}{2}-k}\,k!\,
x^{2(-1)^n\alpha-1-2k}e^{-(-1)^nx^2}
\bigl( 1+{\cal O}(x^{-2})\bigr)\, ;\nonumber
\end{eqnarray}
\item{(iii)} Let
\begin{equation}\label{ab2iii}
(-1)^n(2\alpha-\beta)=2l-1,\quad l\in{\Bbb N}\,,\quad
1-l-(-1)^n\frac{\beta}{2}\notin{\Bbb N}\,,
\end{equation}
then the corresponding solution $y(x)$ of the
fourth Painlev\'e equation P4 possesses the following asymptotic
behavior:
\begin{eqnarray}\label{asymptotic2iii}
&&y=e^{i\frac{\pi}{2}n}y_2(e^{-i\frac{\pi}{2}n}x,(-1)^na,b,c_n)=\nonumber
\\[4pt]
&&=-2x-\frac{2\alpha-\frac{\beta}{2}}{x}+
{\cal O}\bigl(\frac{1}{x^3}\bigr)-
\\[4pt]
&&-\frac{s_{1+n}s_{4+n}}{\pi^{3/2}}\,
e^{i\frac{\pi}{2}(n+1)(-1)^n\beta+i\frac{\pi}{2}(1-n)}\,
\Gamma\bigl(l+(-1)^n\frac{\beta}{2}\bigr)\times\nonumber
\\[4pt]
&&\times 2^{-(-1)^n\frac{\beta}{2}-2l}\,(l-1)!\,
x^{-(-1)^n\beta+2-4l}e^{-(-1)^nx^2}
\bigl( 1+{\cal O}(x^{-2})\bigr)\, ;\nonumber
\end{eqnarray}
\item{(iv)} Let
\begin{equation}\label{ab2iv}
(-1)^n(2\alpha-\beta)=1-2l,\quad l\in{\Bbb N}\,,\quad
-(-1)^n\frac{\beta}{2}\notin{\Bbb N}\,,
\end{equation}
then the corresponding solution $y(x)$ of the
fourth Painlev\'e equation P4 possesses the following asymptotic
behavior:
\begin{eqnarray}\label{asymptotics2iv}
&&y=e^{i\frac{\pi}{2}n}y_+(e^{-i\frac{\pi}{2}n}x,(-1)^na,(-1)^n\beta,c_n)=
\nonumber
\\[4pt]
&&=\frac{\beta}{2x}+\frac{\beta(2\alpha-\frac{3\beta}{2})}{4x^3}+
{\cal O}\bigl(\frac{1}{x^5}\bigr)+
\\[4pt]
&&\!\!\!\!
+(-1)^{l(n-1)}\frac{s_{1+n}s_{4+n}}{\pi^{3/2}}\,
e^{i\frac{\pi}{2}(n+1)(-1)^n\beta+i\frac{\pi}{2}(n-1)}\,
\Gamma\bigl(1+(-1)^n\frac{\beta}{2}\bigr)\!\times\nonumber
\\[4pt]
&&\times 2^{-(-1)^n\frac{\beta}{2}-1-l}\,(l-1)!\,
x^{-(-1)^n\beta-2l}e^{-(-1)^nx^2}
\bigl( 1+{\cal O}(x^{-2})\bigr)\, .\nonumber
\end{eqnarray}
\end{description}
\end{theorem}
{\it Proof}. Application of any one of the B\"acklund transformations to the
genuine solution of P4 generates another genuine solution of P4 with the
different parameters $a$, $b$, or, equivalently, $\alpha$, $\beta$. For $n=0$,
applying (\ref{R+8_action})--(\ref{R-0_action}) to the ``seed" solution $y_0$
(\ref{IK_formula}), we check the resulting asymptotic solutions are of the
type $y_2$, $y_+$ given by (\ref{asymptotic2iii}) and (\ref{asymptotics2iv}).
Applying the same B\"acklund transformations to the asymptotic formulae
(\ref{asymptotics2i})--(\ref{asymptotics2iv}), we see they transform into each
other. The last gives the assertion of the theorem for $n=0$, i.e.\ for the
sector $\arg x\in\bigl(-\frac{\pi}{4};\frac{\pi}{4}\bigr)$. To complete the
proof, it is enough to use the rotation symmetry (\ref{y_symmetry_rotation}).

\smallskip
The cases excluded from Theorems~\ref{Theorem1} and~\ref{Theorem2} above
correspond to the linear submanifolds of the monodromy surface
(\ref{monodromy_surface}), i.e.\ to the classical solutions of P4 described
above.

To complete the description of the class of the degenerated solutions, let
us present the assertions concerning ``less degenerated" solutions from the
article \cite{Kap}.

\begin{theorem}\label{Theorem7} If $x\to\infty$,
$\arg x\in\bigl(-\frac{\pi}{4}+\frac{\pi}{2}n;
\frac{\pi}{4}+\frac{\pi}{2}n\bigr)$ for some $n\in{\Bbb Z}\,$, and
\begin{equation}\label{trivial_sk7}
s_{2+n}=0\,,\quad
s_{1+n}+s_{3+n}\not =0,
\end{equation}
then
\begin{eqnarray}\label{asymptotics7}
&&y(x)=
e^{i\frac{\pi}{2}n}y_+\bigl(e^{-i\frac{\pi}{2}n}x,(-1)^na,(-1)^n\beta,c)=
\\[4pt]
&&=\frac{\beta}{2x}+{\cal O}\bigl(x^{-3}\bigr)+
ax^{-1+2(-1)^n(\alpha-\beta)}e^{-(-1)^nx^2}
\Bigl(1+{\cal O}\bigl(x^{-2}\bigr)\Bigr) ,\nonumber
\end{eqnarray}
where
\begin{eqnarray}
&&a=(-1)^ne^{-i\pi n(-1)^n(\alpha-\beta)}\,
\frac{2^{\frac{1}{2}+(-1)^n(\alpha-\beta)}\sqrt{\pi}}
{\Gamma\bigl(\frac{1}{2}+(-1)^n(\alpha-\frac{\beta}{2})\bigr)
\Gamma\bigl(-(-1)^n\frac{\beta}{2}\bigr)}\times\nonumber
\\[4pt]
&&\times\frac{i}{s_{1+n}+s_{3+n}}
\Bigl(\Theta\bigl(\frac{\pi}{2}n-\arg x\bigr)s_{1+n}-
\Theta\bigl(\arg x-\frac{\pi}{2}n\bigr)s_{3+n}\Bigr).\nonumber
\end{eqnarray}
\end{theorem}

Before the next theorem will be formulated, it is necessary
to note that the assumption
\begin{equation}\label{triviality_assumption}
s_{1+n}+s_{3+n}+s_{1+n}s_{2+n}s_{3+n}=0
\end{equation}
with the equation of the monodromy surface (\ref{monodromy_surface}) together
implies that
\begin{equation}\label{seq1}
1+s_{2+n}s_{3+n}=e^{i\pi (-1)^n\beta }
\end{equation}
or
\begin{equation}\label{seq2}
1+s_{2+n}s_{3+n}=-e^{-i\pi (-1)^n(2\alpha -\beta )}
\end{equation}

\begin{theorem}\label{Theorem8} If $x\to\infty$,
$\arg x\in\bigl(-\frac{\pi}{4}+\frac{\pi}{2}n;
\frac{\pi}{4}+\frac{\pi}{2}n\bigr)$ for some
$n\in{\Bbb Z}\,$, the equations (\ref{triviality_assumption}), (\ref{seq1})
hold, and
\begin{equation}\label{condition8}
s_{2+n}\neq0,\quad \frac{1}{2}+(-1)^n\alpha\notin{\Bbb N}\,,
\end{equation}
then
\begin{eqnarray}\label{asymptotics8}
&&y=e^{i\frac{\pi}{2}n}
y_-\bigl(e^{-i\frac{\pi}{2}n}x,(-1)^na,(-1)^n\beta,c_n)=
\\[4pt]
&&-\frac{\beta}{2x}+{\cal O}\bigl(x^{-3}\bigr)+
cx^{-1+2(-1)^n(\alpha+\frac{\beta}{2})}e^{-(-1)^nx^2}
\Bigl(1+{\cal O}\bigl(x^{-2}\bigr)\Bigr),\nonumber
\end{eqnarray}
where
\begin{eqnarray}\label{c8}
&&c=(-1)^ne^{-i\pi n(-1)^n(\alpha+\frac{\beta}{2})}
2^{-\frac{1}{2}+(-1)^n(\alpha+\frac{\beta}{2})}
i\frac{\Gamma\bigl(\frac{1}{2}-(-1)^n\alpha\bigr)}{
\sqrt{\pi}\,\Gamma\bigl((-1)^n\frac{\beta}{2}\bigr)}\times
\\[4pt]
&&\times\frac{1}{s_{2+n}}\Bigl\{
\Theta\bigl(\frac{\pi}{2}n-\arg x\bigr)
e^{i\pi(-1)^n\alpha}\bigl(s_n+s_{2+n}+s_ns_{1+n}s_{2+n}\bigr)-\nonumber
\\[4pt]
&&-\Theta\bigl(\arg x-\frac{\pi}{2}n\bigr)
e^{-i\pi(-1)^n\alpha}\bigl(s_{2+n}+s_{4+n}+s_{2+n}s_{3+n}s_{4+n}\bigr)
\Bigr\}\,.\nonumber
\end{eqnarray}
\end{theorem}

\begin{theorem}\label{Theorem9} If $x\to\infty$, $\arg x\in
\bigl(-\frac{\pi}{4}+\frac{\pi}{2}n;\frac{\pi}{4}+\frac{\pi}{2}n\bigr)$, for
some $n\in{\Bbb Z}\,$, the equations (\ref{triviality_assumption}), (\ref{seq2})
hold, and
\begin{equation}\label{condition9}
s_{2+n}\neq0,\quad {1\over 2}-(-1)^n\alpha\notin {\Bbb N}\,,
\end{equation}
then
\begin{eqnarray}\label{asymptotics9}
&&y=e^{i\frac{\pi}{2}n}
y_2\bigl(e^{-i\frac{\pi}{2}n}x,(-1)^na,(-1)^nb,c_n\bigr)=
\\[4pt]
&&=-2x+{\cal O}\bigl(x^{-1}\bigr)
-cx^{-(-1)^n(4\alpha-\beta)}e^{-(-1)^nx^2}
\Bigl(1+{\cal O}\bigl(x^{-2}\bigr)\Bigr) ,\nonumber
\end{eqnarray}
where
\begin{eqnarray}\label{c9}
&&c=2^{-(-1)^n(2\alpha-\frac{\beta}{2})}
e^{i\pi n(\frac{1}{2}+(-1)^n(2\alpha-\frac{\beta}{2})}
i\frac{\Gamma\Bigl(\frac{1}{2}+(-1)^n\alpha\Bigr)}{
\sqrt{\pi}\, \Gamma\Bigl\{
\frac{1}{2}-(-1)^n\bigl(\alpha-\frac{\beta}{2}\bigr)\Bigr)}\times\nonumber
\\[4pt]
&&\times
\frac{1}{s_{2+n}}\Bigl\{
\Theta\bigl(\frac{\pi}{2}n-\arg x\bigr)
e^{-i\pi(-1)^n\alpha}\bigl(s_n+s_{2+n}+s_ns_{1+n}s_{2+n}\bigr)-\nonumber
\\[4pt]
&&-\Theta\bigl(\arg x-\frac{\pi}{2}n\bigr)
e^{i\pi(-1)^n\alpha}
\bigl(s_{2+n}+s_{4+n}+s_{2+n}s_{3+n}s_{4+n}\bigr)\Bigr\}\,.
\nonumber
\end{eqnarray}
\end{theorem}

Because the equation
\begin{equation}\label{trivial_sk30}
s_{1+n}+s_{3+n}+s_{1+n}s_{2+n}s_{3+n}=0
\end{equation}
with the equation of the monodromy surface (\ref{monodromy_surface}) and the
condition $\alpha-\frac{1}{2}\in{\Bbb Z}\,$ together imply
$$
1+s_{2+n}s_{3+n}=e^{i\pi (-1)^n\beta }\,,\quad
1+s_{1+n}s_{2+n}=e^{-i\pi (-1)^n\beta }\,,\quad
s_{3+n}+s_{1+n}e^{i\pi (-1)^n\beta }=0\,,$$
it is important to formulate assertion for this limiting case of the
Theorems~\ref{Theorem8} and~\ref{Theorem9}.

\begin{theorem}\label{Theorem3} Let $x\to \infty$,
$\arg x\in\bigl(-\frac{\pi}{4}+\frac{\pi}{2}\,n;
\frac{\pi}{4}+\frac{\pi}{2}\,n\bigr)$ for some $n\in {\Bbb Z}\,$,
and
\begin{equation}\label{trivial_sk3}
s_{2+n}\neq0,\quad
s_{1+n}+s_{3+n}+s_{1+n}s_{2+n}s_{3+n}=0;
\end{equation}
\begin{description}
\item{(i)} if
\begin{equation}\label{ab3i}
(-1)^n\alpha=\frac{1}{2}-k,\quad k\in{\Bbb N}\,,
\end{equation}
then
\begin{eqnarray}\label{asymptotics3i}
&&y=e^{i\frac{\pi}{2}n}y_-(e^{-i\frac{\pi}{2}n}x,(-1)^na,(-1)^n\beta,c_n)=
-\frac{\beta}{2x}+{\cal O}\bigl(\frac{1}{x^3}\bigr)-\nonumber
\\[4pt]
&&-e^{i\frac{\pi}{2}n}(-1)^{k(1-n)}
\Bigl(\frac{s_{4+n}}{s_{2+n}}+e^{-i\pi(-1)^n\beta}\Bigr)
\frac{e^{-i\frac{\pi}{2}(-1)^n\beta(n-2)}}{\sqrt\pi\,
\Gamma\bigl((-1)^n\frac{\beta}{2}\bigr)}\times\nonumber
\\[4pt]
&&\times 2^{-k+(-1)^n\frac{\beta}{2}}\,(k-1)!\,
x^{-2k+(-1)^n\beta}e^{-(-1)^nx^2}
\Bigl(1+{\cal O}\bigl(\frac{1}{x^2}\bigr)\Bigr)\,;\nonumber
\end{eqnarray}
\item{(ii)} if
\begin{equation}\label{ab3ii}
(-1)^n\alpha=-\frac{1}{2}+l,\quad l\in{\Bbb N}\,,
\end{equation}
then
\begin{eqnarray}\label{asymptotics3ii}
&&y=e^{i\frac{\pi}{2}n}y_2(e^{-i\frac{\pi}{2}n}x,(-1)^na,b,c_n)=
-2x+{\cal O}\bigl(\frac{1}{x}\bigr)+
\\[4pt]
&&+e^{-i\frac{\pi}{2}n}(-1)^l
\Bigl(\frac{s_{4+n}}{s_{2+n}}+e^{-i\pi(-1)^n\beta}\Bigr)
\frac{e^{-i\frac{\pi}{2}(-1)^n\beta(n-2)}}{\sqrt\pi\,
\Gamma\bigl((-1)^n\frac{\beta}{2}+1-l\bigr)}\times\nonumber
\\[4pt]
&&\times 2^{-2l+1+(-1)^n\frac{\beta}{2}}\,(l-1)!\,
x^{-4l+2+(-1)^n\beta}e^{-(-1)^nx^2}
\Bigl(1+{\cal O}\bigl(\frac{1}{x^2}\bigr)\Bigr)\,.\nonumber
\end{eqnarray}
\end{description}
\end{theorem}

\vskip 2em
{\bf Acknowledgments.} The author thanks the Russian Foundation for
Fundamental Investigations for its support under grant number 96-01-00668.

\newpage
\ifx\undefined\bysame
\newcommand{\bysame}{\leavevmode\hbox to3em{\hrulefill}\,}
\fi

\end{document}